\documentclass[letter,11pt]{article}
 
\usepackage{jheppub}

\usepackage{tcolorbox}
\usepackage{hyperref}
\usepackage{graphicx}
\usepackage{amsmath}
\usepackage{amssymb} 
\usepackage{xspace}
\usepackage{slashed}
\usepackage{subcaption}
\usepackage[normalem]{ulem}
\usepackage[utf8]{inputenc}
\usepackage[T1]{fontenc}
\usepackage{mathtools}
\usepackage{stmaryrd}
\usepackage{enumerate}
\usepackage{mathrsfs}

\newcommand{\nocontentsline}[3]{}
\newcommand{\tocless}[2]{\bgroup\let\addcontentsline=\nocontentsline#1{#2}\egroup}

\newcommand{\nn}{\nonumber}

\newcommand{\bea}{\begin{eqnarray}}
\newcommand{\eea}{\end{eqnarray}}

\def\ln{\textrm{ln}}

\def\nn{\nonumber}

\allowdisplaybreaks[2]

\usepackage{marginnote}

\preprint{\begin{flushright}
MIT--CTP 5327
\end{flushright}}

\title{Probing a dilute short lived Quark Gluon Plasma medium with jets}

\author{Varun Vaidya}
\affiliation{Center for Theoretical Physics, Massachusetts Institute of Technology, Cambridge, MA~02139, U.S.A.}

\emailAdd{vvaidya@mit.edu}

\abstract{I look at  propagation of jets that act as a probe of the medium in the phenomenologically relevant case of a short lived dilute Quark Gluon Plasma(QGP) created during heavy ion collisions.  Working in the regime where  the lifetime of the medium is of the order or smaller  than the formation time of the energetic jet,  I derive a factorization formula with a manifest separation of scales for an illustrative jet substructure observable using the Open Quantum system Effective Field Theory(EFT) approach developed in \cite{Vaidya:2020lih}, in terms of a medium structure function and a medium induced jet function .  The  medium structure function or the "PDF of the medium" that describes the observable independent physics of the QGP remains identical to that in a dilute long lived medium considered in \cite{Vaidya:2021vxu},  while the medium induced jet function is modified and incorporates quantum interference between the hard interaction that creates the jet and subsequent medium interactions.  The medium jet function continues to enjoy a BFKL rapidity evolution just as for a long lived medium,  albeit with a modified renormalization scale that now depends on the medium lifetime. 
}

\begin{document}
\maketitle

\section{Introduction}

The high energy collision of nuclei both at RHIC and the LHC creates sufficiently energetic partons that can escape confinement from color neutral hadrons and give rise to a strongly/weakly coupled soup of deconfined quarks and gluons known as the Quark Gluon Plasma(QGP) medium. Experimental evidence suggests that this medium behaves as a near perfect liquid in thermal equilibrium with very low viscosity but exists for a very short time ($\sim 10$ fm/c). At very short distances,  we can think of this plasma as consisting of quasi free quarks and gluons with typical energy of the order of the temperature(T) of the medium which is usually much lower than the center of mass energy of the initiating nuclear collision. The stopping nuclear collisions which create the QGP are accompanied by hard interactions which create highly energetic partons (E $\gg$ T) which eventually form jets. These jets then have to traverse through a region of the hot QGP as they evolve and hence they get modified in heavy ion collision, compared with proton-proton collisions where the jets evolve in a vacuum background.
The modification of a jet in a medium compared to its vacuum evolution can tell us something about the properties of the medium, making them useful hard probes of the Quark Gluon Plasma. A phenomenon that has been extensively studied in  literature\cite{Gyulassy:1993hr,Wang:1994fx,Baier:1994bd,Baier:1996kr,Baier:1996sk,Zakharov:1996fv,Zakharov:1997uu,Gyulassy:1999zd,Gyulassy:2000er,Wiedemann:2000za,Guo:2000nz,Wang:2001ifa,Arnold:2002ja,Arnold:2002zm,Salgado:2003gb,Armesto:2003jh,Majumder:2006wi,Majumder:2007zh,Neufeld:2008fi,Neufeld:2009ep} is that of Jet quenching, which involves a suppression of particles with high transverse momenta in the medium. This has also been recently observed in experiments at both Relativistic Heavy Ion Collider (RHIC) \cite{Arsene:2004fa,Back:2004je,Adams:2005dq,Adcox:2004mh} and Large Hadron Collider (LHC) \cite{Aad:2010bu,Aamodt:2010jd,Chatrchyan:2011sx}. The suppression mechanism happens through the mechanism of energy loss which can happen either through a collision of the energetic partons in the jet with the soft partons of the medium or through medium-induced radiation, but the latter dominates at high energy. The key to understand jet quenching and jet substructure modifications in heavy ion collisions is to understand how the jet interacts with the expanding medium. There has been extensive theoretical effort to study the jet energy loss mechanism (see Refs.~\cite{Mehtar-Tani:2013pia,Blaizot:2015lma,Qin:2015srf,Cao:2020wlm} for recent reviews). All these attempts rely on a direct Feynman diagram calculation and the information about the medium is encoded in the form of a transport co-efficient $\hat q$ which measures the average transverse momentum gained by a parton per unit time.

Given that the evolution of the jet in the medium is a multi-scale problem with a hierarchy of scales,  a systematic Effective Field Theory(EFT) treatment which allows us to separate the physics of the medium from that of the jet at the level of gauge invariant operators is currently missing in literature. At the same time, a complete analytical calculation which includes the hard interaction which creates the jet along with the subsequent medium interactions is also necessary for comparison with data. 
In \cite{Vaidya:2020lih}, I proposed an EFT approach using tools of Open Quantum systems and SCET(Soft Collinear Effective Theory)(\cite{Bauer:2002aj,Bauer:2003mga,Manohar:2006nz,Bauer:2000yr,Bauer:2001ct,Bauer:2002nz}) for jet substructure observables in heavy ion collisions. This involved a factorization formula for an illustrative jet substructure observable in terms of matrix elements of gauge invariant operators. The advantages of using an EFT approach were also clearly outlined in that paper, some of them being a manifest separation of physics at widely separated scales along with a gauge invariant operator definition for the physics at each scale. In particular, this led to a universal parton distribution for the QGP medium which encodes the physics of the QGP in an observable independent manner. 

The evolution of the jet in the medium usually depends on multiple scales which can be broadly divided into three categories: Kinematic scales such as such as the jet energy, thermal scales of the QGP , namely the temperature T and the Debye screening mass $m_D \sim gT$ and the size or equivalently the temporal extent of the medium. The second class of scales that appear due to dynamical evolution of the system are the mean free path of the jet ($\lambda_{mfp}$) which depends on the density of the medium and the strength of the jet-medium interaction,  and the formation time of the jet partons given in terms of their energy(Q) and transverse momentum ($q_T$) as $t_f\sim Q/q_T^2$ , which is the time scale over which the partons in the jet maintain their coherence. The third and final category are the measurement scales imposed on the final state jet.
The hierarchy between these scales can vary widely depending on the details of the experiment. However, in general the jet energy scale(Q) will be a hard scale much larger than all the other scales. For sufficiently high temperature(T $\gg \Lambda_{QCD}$), we can assume $g \ll 1$ and so that $m_D \ll T$ which is what we will assume for the rest of this paper. In current heavy ion collision experiments, the temperature achieved lies in the range $150 - 500$ MeV, and may not always be a perturbative scale. Thus, a fully weak coupling calculation may not be valid. In this paper, I will stick to the case of a weakly coupled QGP for simplicity. While this will enable me to do a perturbative calculation, it is not a requirement for deriving a factorization of the physics at widely separated scales. For $T \sim \Lambda_{QCD}$, some of the functions in the factorization formula become non-perturbative and would then need to be extracted from lattice/experiment.

  A first simple case  of a long-lived dilute medium with the medium length(L) and the mean free path($\lambda_{\text{mfp}}$) much larger than all other time scales in the problem was investigated in \cite{Vaidya:2021vxu}.  A factorization formula for an illustrative jet substructure observable was derived showing a manifest separation of hard and soft scales and for the first time, providing a gauge invariant operator definition for a Parton Distribution function of the medium. It was also shown that the PDF of the medium evolves in rapidity with the BFKL equation.
  
  However, the assumed hierarchy of time scales may not always be a realistic scenario, in particular the formation time of the jet($t_F$) would be expected to be much larger than L in current experiments especially for high energy jets. At the same time, a dense medium will lead to a small value for $\lambda_{\text{mfp}}$ in which case the well known LPM(Landau-Pomeranchuk-Migdal) effect (\cite{CaronHuot:2010bp,Ke:2018jem,Mehtar-Tani:2019ygg,Arnold:2015qya,Arnold:2016kek}) will become important.  
  
  In this paper, I take the next step towards this phenomenologically relevant case, by working in a regime where $L \leq t_f$ while maintaining $ \lambda_{\text{mfp}} \gg t_f$ which can be thought of as a short lived dilute medium. Using the same tools of Soft-Collinear Effective Theory(SCET) and Open quantum systems, employed for the case of the dilute long lived medium in \cite{Vaidya:2020lih},  I will derive a factorization formula for a generic jet substructure observable.  This shows the power of an EFT approach since a manifest separation of physics at separated scales shows that the PDF of medium is unaffected by the medium length and is identical to the case of a long lived medium, which therefore continues to enjoy a BFKL evolution. The medium induced jet function is now modified but continues to have the same renormalization group equation, but with an altered natural scale.  The jet function now incorporates quantum interference between the hard interaction that created the jet and the subsequent forward scattering interactions with the medium.  The case of a dense medium will be considered in a future work\cite{varun}.

This paper is organized as follows.
In Section~\ref{sec:Obs}, I review the details of the observable and 
the hierachy of scales. Section \ref{sec:Fact} looks in detail at the derivation of a factorization formula for the case of a short lived medium. In Section \ref{sec:Loop}, I detail the one loop corrections from medium induced Bremsstrahlung for the medium induced jet function which also gives their Renormalization Group(RG) equation. Section~\ref{sec:Resum} looks at the resummation of large logarithms by solving RG equations and looks at possible extension of the formalism for a dense medium. A summary and analysis of results along with future directions in provided in Section~\ref{sec:Conclusion}. Details of the loop calculations are provided in the Appendix.


\section{ Jet substructure in Heavy Ion collisions}
\label{sec:Obs}
In \cite{Vaidya:2020lih},  I developed an EFT formalism for jet substructure observables in a heavy ion collision environment, writing down a factorization formula for an illustrative observable: The transverse momentum imbalance between groomed dijets along with a cumulative jet mass measurement on each jet. This observable was introduced to allow for a clean measurement while countering the issue of jet selection bias. We want to consider final state fat (jet radius R $\sim $ 1) dijet events produced in a heavy ion collision in the background of a QGP medium. The jets are isolated using a suitable jet algorithm such as anti-kT with jet radius $R\sim 1$. We examine the scenario when the hard interaction creating the back to back jets happens at the periphery of the heavy ion collision so that effectively only one jet passes through the medium while the other evolves purely in vacuum. \\
Since, it is hard to give a reliable theory prediction for the distribution of soft hadrons from the cooling QGP medium, we put a grooming on the jets.  We employ the soft drop grooming algorithm \cite{Larkoski:2014wba} with $\beta=0$ and an energy cut-off sufficiently large to remove all partons at energy T and lower.  Given a hard scale Q $\sim 2E_J$, where $E_J$ is the energy of the jet and an energy cut-off, $z_{c}E_J$, we work in the hierarchy 
\bea
Q \sim z_{c}Q \gg T 
\eea
where T is the plasma temperature.  The measurement we wish to impose is the transverse momentum imbalance between the two jets and we want to to give predictions for the regime $q_T \sim T$. We are going to assume a high temperature weak coupling $g \ll 1$ scenario so that the Debye screening mass $m_D$ is parametrically much smaller than the temperature.

While this constrains the radiation that lies outside the groomed jet, we still need to ensure that the radiation that passes grooming and hence forms the jet has a single hard subjet to counter jet selection bias as was explained in \cite{Vaidya:2020lih}. This can be ensured by putting a cumulative jet mass measurement $e_i$ on each groomed jet with the scaling $e_i \sim  T^2/Q^2 $.  

While I am choosing a specific jet substructure observable, the factorization formula that I will derive is quite generic and can be easily modified to suit any observable. The factorization formula will be written in terms of the vacuum cross section, a medium structure function and a medium induced jet function.  For any observable, the form of the factorization  will remain unchanged as will the definition of the medium structure function which describes the observable independent physics of the medium. At the same time, by Renormalization group consistency, the anomalous dimension of the medium jet function will also remain the same but the scale of renormalization and any finite corrections will change depending on the observable.
 
We wish to write down a factorization theorem  which separates out functions depending on their scaling in momentum space with $ \lambda = q_T/Q \sim T/Q \sim \sqrt{e_i}$ as the expansion parameter of our EFT. The dominant interaction of the jet with the medium involves forward scattering of the jet in the medium and is mediated by the Glauber mode.  Using an open quantum systems approach combined with the EFT for forward scattering developed in \cite{Rothstein:2016bsq}, a factorization formula for the reduced density matrix of the jet was derived in \cite{Vaidya:2020lih}. This factorization was derived under the assumption of a long lived dilute medium meaning
\begin{itemize}
\item
The lifetime of the medium $t_M$ which is equivalent to the time of propagation of the jet in the medium and the mean free path $\lambda_{\text{mfp}}$ of the jet are larger than than the formation time of the jet $t_f$.
\item
The medium is translationally invariant.
\end{itemize}
These assumptions implied that the dominant contribution to the cross section was when the partons created in the jet went on-shell before interacting with the medium and successive interactions of the jet with the medium were incoherent. 

While this was a good first step, a more realistic scenario is when the lifetime of the medium is of the order or smaller than the formation time($t_f$) of the energetic jet partons.  For eg,  for a temperature T of the order of a few GeV,  and a jet with energy of a few tens to hundreds of GeV,   the formation time for the most energetic of  partons , $t_f^c \sim E_{\text{jet}}/q_T^2 \sim E_{\text{jet}}/T^2  $ will range from a few fermi to a hundreds of fermi.   On the other hand,  the typical lifetime of the QGP medium ($t_M$) is of the order of 10 fermi. This compels us to look at two possible  
scale hierarchies in our EFT formulation $ t_f^c \sim t_M$ and $t_M \ll  t_f^c$. At the same time  formation time of soft radiation $t_f^s \sim 1/T$ so that we can safely work in the regime $t_f^s \ll t_M$.  This means that the soft radiation effectively sees an infinite medium. 

We will continue to work in the regime of a dilute medium , i.e., the mean free path of the jet $\lambda_{\text{mfp}}$  is much larger than the formation time $t_f^c$.  The more involved case of a dense medium will be addressed in a companion paper \cite{varun}. 
\section{Factorization for Reduced density matrix in a short lived medium}
\label{sec:Fact}
The objective is to derive a factorization formula for the observable described in the previous section showing a clear separation of scales in terms of gauge invariant matrix elements.  I will follow the procedure outlined in \cite{Vaidya:2020lih}, treating the jet as an open quantum system interacting with the QGP bath and follow the time evolution of the reduced density matrix of the jet.  Along the way, I will work out the factorization for this density matrix.  I will work within the framework of Soft Collinear Effective Theory(SCET) which has been extensively used to rigorously derive factorization formulas for high energy scattering experiments.  This EFT was extended to include the regime of forward scattering mediated by the so called Glauber modes in \cite{Rothstein:2016bsq}.  This is ideally suited for my purpose since the interaction of the jet with the medium is dominated by forward scattering.

For ease of analysis, we consider that the hard interaction that creates the jet is an $e^+e^-$ collision. While this is not a real scenario, it is an ideal playground to work out the EFT framework which mainly deals with the final state physics. The EFT structure can then be easily carried over to the realistic case of nuclear/hadronic collisions, which we leave for future analysis as part of a detailed phenomenological application of the formalism developed in this paper. 

As explained in \cite{Vaidya:2020lih},  we can choose a convenient light-like directions $n \equiv (1,  0,  0,  1)$ and $\bar n \equiv (1,  0,  0, - 1)$ for the initiating back to back $q\bar q$ pair that eventually forms our dijet.  Without loss of generality,  we assume the n jet traverses the medium while the $\bar n$ jet evolves in vacuum. The energetic partons in the n  jet then scale in momentum space light-cone co-ordinates as $Q(1, \lambda^2,  \lambda)$  and form the n $collinear$ degrees of freedom in our EFT. 
The QGP medium consists of  partons which scale in momentum space as $Q(\lambda, \lambda, \lambda)$ and form the $soft$ degrees of freedom.  

The hard interaction that creates the initiating $q\bar q$ pair can be encoded using an effective current operator 
\bea
\mathcal{O}_{H} =  C(Q)L_{\mu} J_{SCET}^{\mu} 
\eea
$C(Q)$ is the Wilson co-efficient for this contact operator that depends only on the hard scale Q. This will lead to a hard function H(Q) at the amplitude squared level and its form is discussed in \cite{Vaidya:2020lih}.  $L^{\mu}$ is the initial state current, which in this case is just the lepton current, while $J^{SCET}_{\mu} $ would be the final state SCET current, which is just the gauge invariant quark current. 
\bea
L^{\mu} = \bar l \gamma^{\mu} l, \ \ \  J_{SCET}^{\mu}= \bar \chi_n \gamma^{\mu} \chi_{\bar n} \nn
\eea

 The initial state density matrix would then be 
\bea
\rho(0) = |e^+ e^-\rangle \langle e^+e^-| \otimes \rho_B
\eea 
 We have started with the assumption that the initial state participating in the hard interaction is disentangled from the state of the background medium.

We want to consider a scenario where the medium exists for a finite time, which is of the order of the formation time of the jet.  Since we are mainly interested in the impact of this constraint on the evolution of the  jet in the medium,  we can cast the problem in terms of a  finite time of interaction of the jet with the medium.  Since the jet-medium interaction is mediated by effective forward scattering operators,  this effectively constrains these operators to be relevant for only a finite time during jet evolution.

 We can follow the evolution of this density matrix which will evolve with the effective Hamiltonian 
\bea
H=H^{\text{IR}}+ \mathcal{O}_{H}
\eea
We are going to write 
\bea
H^{\text{IR}} = H^{SCET} + H^G(t) = H_n+H_s+H^G\Theta(t_M-t)
\eea
We have put in an explicit time dependence in the Glauber Hamiltonian which turns it off at t=$t_M$ when the jet-medium interaction ceases.
$H_n$ and $H_s$  describes the evolution of the collinear and soft degrees of freedom.  $H^G$ describes the forward interaction between the soft and collinear degrees of freedom. 

We follow the same series of steps as in \cite{Vaidya:2020lih}, now with an explicit time dependent interaction Hamiltonian.
The time evolved density matrix is given as 
\bea
\rho(t)&=& \mathcal{T}\Big\{e^{-i\int_0^t dt'H(t')}\Big\}\rho(0)\bar{\mathcal{T}}\Big\{e^{i\int_0^t dt'H(t')}\Big\} 
\eea
where the time evolution is enforced by a time ordered product due to the explicit time dependence of the Hamiltonian.  We can now insert an identity on both sides with a view of first expanding to leading order in $\mathcal{O}_H$,
\bea
\rho(t)&=&\Bigg[\mathcal{T}\Big\{e^{-i\int_0^t dt'H^{\text{IR}}(t')}\Big\}  \bar{\mathcal{T}}\Big\{e^{i\int_0^t dt'H^{\text{IR}}(t')}\Big\}\Bigg]\mathcal{T}\Big\{e^{-i\int_0^t dt'H(t')}\Big\}\rho(0)\bar{\mathcal{T}}\Big\{e^{i\int_0^t dt'H(t')}\Big\}\nn\\
&&\Bigg[\mathcal{T}\Big\{e^{i\int_0^t dt'H^{\text{IR}}(t')}\Big\}\bar{\mathcal{T}}\Big\{e^{i\int_0^t dt'H^{\text{IR}}(t')}\Big\}\Bigg] \nn\\
&=& \mathcal{T}\Big\{e^{-i\int_0^t dt'H^{\text{IR}}(t')}\Big\}U(t,0)\rho(0)U(0,t) \bar{\mathcal{T}}\Big\{e^{i\int_0^t dt'H^{\text{IR}}(t')}\Big\}
\eea
where the evolution operator is defined as 
\bea
 U(t,0)=\bar{\mathcal{T}}\Big\{e^{i\int_0^t dt'H^{\text{IR}}(t')}\Big\}\mathcal{T}\Big\{e^{-i\int_0^t dt'H(t')}\Big\}
\eea
Our evolution operator $U(t,0)$ now obeys the equation 
\bea
\partial_t U(t,0)= -i\mathcal{O}_{H,\text{IR}}(t)U(t,0),  \ \ \  \text{with} \ \ \  \mathcal{O}_{H,\text{IR}}(t)=\bar{\mathcal{T}}\Big\{e^{i\int_0^t dt'H^{\text{IR}}(t')}\Big\}\mathcal{O}_{H}\mathcal{T}\Big\{e^{-i\int_0^t dt'H^{\text{IR}}(t')}\Big\}
\eea
which has the solution 
\bea
U(t,0) = \mathcal{T} \Big\{ e^{-i\int_0^t dt' \mathcal{O}_{H,\text{IR}}(t')} \Big\} 
\eea

which is the evolution operator written as a time ordered exponent of the dressed hard operator. 
Our solution for the density matrix now becomes 
\bea
 \rho(t)=  \mathcal{T}\Big\{e^{-i\int_0^t dt'H^{\text{IR}}(t')}\Big\}\mathcal{T} \Big\{ e^{-i\int_0^t dt' \mathcal{O}_{H,\text{IR}}(t')} \Big\} \rho(0) \bar{\mathcal{T}} \Big\{ e^{-i\int_0^t dt' \mathcal{O}_{H,\text{IR}}(t')} \Big\} \bar{\mathcal{T}}\Big\{e^{i\int_0^t dt'H^{\text{IR}}(t')}\Big\}
\eea
We are interested in creating dijets, it is sufficient to consider a single insertion of the Hard operator on each side of the cut. 
\bea
 \rho(t)&=& \mathcal{T}\Big\{e^{-i\int_0^t dt'H^{\text{IR}}(t')}\Big\}\rho(0)\bar{\mathcal{T}}\Big\{e^{i\int_0^t dt'H^{\text{IR}}(t')}\Big\}\nn\\
&+& \int_0^t dt_1\int_0^t dt_2\mathcal{T}\Big\{e^{-i\int_0^t dt'H^{\text{IR}}(t')}\Big\}\mathcal{O}_{H,\text{IR}}(t_1)\rho(0) \mathcal{O}^{\dagger}_{H,\text{IR}}(t_2)\bar{\mathcal{T}}\Big\{e^{i\int_0^t dt'H^{\text{IR}}(t')}\Big\}
\eea
When we impose the measurement for the dijet event with the required properties, only the second term will survive, hence hereafter we can simply follow the evolution for this piece.

We define 
\bea
\sigma(t) &=&   \int_0^t dt_1\int_0^t dt_2 \mathcal{T}\Big\{e^{-i\int_0^t dt'H^{\text{IR}}(t')}\Big\}\mathcal{O}_{H,\text{IR}}(t_1)\rho(0) \mathcal{O}^{\dagger}_{H,\text{IR}}(t_2)\bar{\mathcal{T}}\Big\{e^{i\int_0^t dt'H^{\text{IR}}(t')}\Big\}\nn\\
&=& |C(Q)|^2I_{\mu\nu}\int d^3x_1\int_0^t dt_1\int d^3x_2\int_0^t dt_2 e^{-i(x_1-x_2)\cdot(p_e+p_{\bar e})}\nn\\
&&\mathcal{T}\Big\{e^{-i\int_0^t dt'H^{\text{IR}}(t')}\Big\}J^{\mu}_{SCET,\text{IR}}(x_1)|0\rangle \langle 0| J^{\nu}_{SCET,\text{IR}}(x_2)\bar{\mathcal{T}}\Big\{e^{i\int_0^t dt'H^{\text{IR}}(t')}\Big\}
\eea
where $p_e$, $p_{\bar e}$ are the momenta of the initial state electron positron. In the c.o.m. frame $p_e+p_{\bar e} =(Q, 0, 0, 0)$.$I^{\mu \nu}$ is the Lepton tensor. The Glauber Hamiltonian is expressed in terms of effective gauge invariant operators for quark-quark ($qq$), quark-gluon ($qg$ or $gq$) and gluon-gluon ($gg$) interactions which have been worked out in the Feynman gauge in Ref.~\cite{Rothstein:2016bsq}
\bea
\label{EFTOp}
H^G &=& \sum_{ij} C_{ij}\mathcal{O}_{ns}^{ij} \nn\\
\mathcal{O}_{ns}^{qq}&=&\mathcal{O}_n^{qB}\frac{1}{\mathcal{P}_{\perp}^2}\mathcal{O}_s^{q_nB} , \ \ \ \mathcal{O}_{ns}^{qg}=\mathcal{O}_n^{qB}\frac{1}{\mathcal{P}_{\perp}^2}\mathcal{O}_s^{g_nB} , \nn\\
\mathcal{O}_{ns}^{gq}&=&\mathcal{O}_n^{gB}\frac{1}{\mathcal{P}_{\perp}^2}\mathcal{O}_s^{q_nB} , \  \ \  \mathcal{O}_{ns}^{gg}=\mathcal{O}_n^{gB}\frac{1}{\mathcal{P}_{\perp}^2}\mathcal{O}_s^{g_nB}
\eea
where $B$ is the color index and the subscripts $n$ and $s$ denote the collinear and soft operators. The Glauber gluon propagator appears as the derivative in $\perp$ direction. We will assume for the remainder of the paper that the jet which traverses the medium points along the n direction. $C_{ij}$ are the Wilson co-efficients for these contact operators and all begin at $O(\alpha_s)$.

The SCET current is given as 
\bea
J^{\mu}_{SCET}= \Big[S_n^{\dagger}S_{\bar n}\Big]\bar \chi_n W_n^{\dagger} \gamma^{\mu} W_{\bar n}\chi_{\bar n}
\label{JSCET}
\eea

where $S_i, W_j$ are soft and collinear Wilson lines, defined as 
\bea
S_n^{(r)}(x) = \text{P} \exp \Bigg[ig \int_{-\infty}^0 ds n \cdot A_s^B(x+sn)T^B_{(r)}\Bigg]\nn\\
 W_n(x)=\Bigg[\sum_{\text{perms}} exp\left(-\frac{g}{n\cdot\mathcal{P}}\bar n\cdot A_n(x)\right)\Bigg]
\eea
We see that the SCET current is already factorized in terms of Soft and Collinear sectors which are manifestly gauge invariant. At the same time they are decoupled from each other in $H^{SCET}$. However, the  Glauber Hamiltonian prevents us from factorizing the full density matrix since it couples the collinear n and the Soft sectors. We therefore need to expand in powers of the Glauber Hamiltonian and establish factorization at each order in the Glauber expansion.  In this paper, we will only work to quadratic order in $H^G$ which is justified for the case of a dilute medium. The case of a dense medium requires us to include all higher order piece and will be the subject of an upcoming paper \cite{varun}.

To proceed, we rearrange the the result above so as to be able to do a systematic expansion in the Glauber Hamiltonian.  This was also outlined in \cite{Vaidya:2020lih} so we will not repeat it here.
For convenience lets define 
\bea
\int d \tilde x =  \int d^3x_1\int_0^t dt_1\int d^3x_2\int_0^t dt_2e^{-i(x_1-x_2)\cdot(p_e+p_{\bar e})}
\eea

\bea
 \sigma(t)&=&  |C(Q)|^2 I_{\mu\nu}\int d \tilde x e^{-iH^{SCET}t}\mathcal{T} \Big\{ e^{-i\int_0^t dt' H^G_{I_{sc}}(t')} J^{\mu}_{SCET,I_{sc}}(x_1)\Big\} |0\rangle \nn\\
&& \langle 0| \bar{\mathcal{T}} \Big\{ e^{-i\int_0^t dt' H^G_{I_{sc}}(t')}J^{\nu}_{SCET,I_{sc}}(x_2) \Big\} e^{iH^{SCET}t}
\label{GExpp}
\eea
where 
\bea
O_{I_{sc}}(t) =  e^{iH^{SCET}t}O(t)e^{-iH^{SCET}t} 
\eea

so that all operators are now dressed with the SCET Hamiltonian. We are now set up to do an expansion in the Glauber Hamiltonian. Ultimately, for this paper, we want to compute the trace over the reduced density matrix with an appropriate measurement 
\bea
\label{Sigm}
\Sigma(t) \equiv \text{Tr}[\sigma(t) \mathcal{M}]\big|_{t\rightarrow \infty} 
\eea
where we first completely trace over the Soft and collinear degrees of freedom with the measurement $\mathcal{M}$ which includes the jet algorithm to isolate a final state large radius groomed dijet configuration with the required $q_T$ imbalance and jet mass. 
We then expand this out in powers of $H^G$
\bea
\label{LEQ}
 \Sigma(t) = \Sigma^{(0)}(t) + \Sigma^{(1)}(t)+ \Sigma^{(2)}(t)+ O([H^G]^3)+...
\eea
In the next section we will sketch the proof for factorization of the reduced density matrix upto quadratic order in the $H^G$ expansion. 

\subsection{Leading order in Glauber:Vacuum evolution}

We start with the leading order term from Eq.\ref{GExpp} with no Glauber insertions and so should simply give us a result proportional to the vacuum-background cross section.  
\bea
\sigma^{(0)}(t) &=&  |C(Q)|^2I_{\mu\nu}\int d \tilde x e^{-iH_{SCET}(t-t_1)} J^{\mu}_{SCET}(\vec{x}_1)e^{-iH^{SCET}t_1} |0\rangle \nn\\
&&\langle 0|e^{iH^{SCET}t_2} J^{\nu}_{SCET}(\vec{x}_2)e^{iH^{SCET}(t-t_2)}\nn
\eea

We put in our measurement on the dijets and take a trace over final states and take the limit $t \rightarrow \infty$.
\bea
&&\langle X| \sigma(t \rightarrow \infty)\mathcal{M}|X\rangle \equiv \Sigma^{(0)}
\eea

where we have 
\bea
\mathcal{M} = \delta^{2}\left( \vec{q}_T- \vec{p}^{\perp}_{n,\in \text{gr}}-\vec{p}^{\perp}_{\bar n, \in \text{gr}}\right)\Theta\left(e_n- e_{n,\in \text{gr}}\right) \Theta \left(e_{\bar n}-e_{\bar n,\in \text{gr}}\right) \nn
\eea
which imposes the transverse momentum  $\vec{q}_T$ and the cumulative jet mass on the groomed jet.
Since the lifetime of the medium has no impact on this piece, the factorization remains exactly same as in \cite{Vaidya:2020lih} so I will only quote the final result here
\bea
\label{SigmZ}
\Sigma^{(0)}(q_T,e_n,e_{\bar n}) =V \times H(Q,\mu)S(\vec{q}_T;\mu,\nu)\otimes_{q_T}\mathcal{J}^{\perp}_n(e_n,Q,z_{c},\vec{q}_T;\mu,\nu)\otimes_{q_T} \mathcal{J}^{\perp}_{\bar n}(e_{\bar n},Q,z_{c},\vec{q}_T;\mu,\nu)\nn\\
\eea
where $\otimes_{q_T}$ indicates a convolution in $\vec{q}_T$.
H(Q) is the hard function which also includes the born level term. V is the 4d volume. The soft function is defined as 
\bea
\label{eq:soft}
 S(\vec{q}_T) = \frac{1}{N_R}\text{tr}  \langle X_S| \mathcal{T}\Big\{e^{-i\int_0^{\infty}dt' H_{S}(t')}S_{\bar n}^{\dagger}S_n(0)\Big\}|0\rangle \langle 0| \mathcal{\bar T}\Big\{e^{-i\int_0^{\infty}dt' H_S(t')}S_{n}^{\dagger}S_{\bar n}(0)\Big\}\delta^2(\vec{q}_T-\mathcal{P}_{\perp})|X_S\rangle \nn
\eea
The trace here is a trace over color and its understood that $|X_S\rangle \langle X_S|$ includes a sum over soft states with their phase space integrated over.  The quark jet function is defined as 
\bea
\mathcal{J}_n^{\perp}(e,Q) &=& \frac{(2\pi)^3}{N_c} \text{tr}\langle X_n| \mathcal{T}\Big\{e^{-i\int_0^{\infty}dt' H_n(t')}\bar \chi_n(0)\Big\}|0\rangle\nn\\
&& \langle 0|\mathcal{\bar T}\Big\{e^{-i\int_0^{\infty}dt' H_n(t')}\frac{\slashed{\bar n}}{2}\chi_n\Big\}\delta(Q-\mathcal{P}^-)\delta^2(\mathcal{P}^{\perp})\Theta(e_n- \mathcal{E}_{\in n,\text{gr}})\delta^2(\vec{q}_T-\mathcal{P}^{\perp}_{\not\in n,\text{gr}})|X_n\rangle \nn
\eea
  The Renormalization group equation for each function can be solved in impact parameter space where the convolution in $\vec{q}_T$ turns into a product 
\bea
 \Sigma^{(0)} =V \times H(Q,\mu)\times \int d^2b e^{i \vec{q}_{T}\cdot \vec{b}}S(\vec{b};\mu,\nu)\mathcal{J}^{\perp}_n(e_n,Q,z_{cut},\vec{b} ;\mu, \nu) \mathcal{J}^{\perp}_{\bar n}(e_{\bar n},Q,z_{cut},\vec{b};\mu,\nu)\nn\\
\eea
where for a function $f(\vec{q}_T)$,
\bea
f(\vec{b}) = \int \frac{d^2\vec{q}_T}{(2\pi)^2}e^{-i \vec{q}_T \cdot \vec{b}}f(\vec{q}_T)
\eea
The one loop results along with RG equations for all the functions in $\vec{b}$ space were presented in Appendix of \cite{Vaidya:2020lih}.

\subsection{Next-to-Leading order in Glauber}

We now consider the next to leading order term in the expansion of the Glauber Hamiltonian starting from Eq.\ref{Sigm}.  As explained in \cite{Vaidya:2020lih} we need to do atleast a quadratic Glauber insertion as we can ignore the interference between the soft emissions off the hard vertex and the soft partons sourced by medium due a small formation time for the soft mode. There will still be quantum interference for the energetic collinear emissions off the hard vertex and those induced by the medium,which we shall include.  
With a quadratic Glauber insertion, we can have two contributions depending on whether the two Glauber insertions are on the same or opposite sides of the cut. This respectively corresponds to a single virtual and real interaction of the jet with the medium.  
\bea
\Sigma^{(2)}(t) =  \Sigma_R^{(2)}(t)+\Big\{\Sigma_V^{(2)}(t)+c.c. \Big\}
\eea

We will look at the factorization in detail to account for the explicit time dependence of the Glauber Hamiltonian.
\subsubsection{ Glauber insertion on both sides of the cut}
Starting with Eq. \ref{GExpp}, we can expand to quadratic order inserting a Glauber vertex on each side of the cut,
\bea
\sigma_{R}^{(2)}(t) &=& |C(Q)|^2I_{\mu \nu}\int d\tilde x e^{-iH^{SCET}t}\mathcal{T}\Big\{\int_0^t dt'H^G_{I_{sc}}(t')J^{\mu}_{SCET,I_{sc}}(x_1)\Big\}|0\rangle \nn\\
&&\langle 0|\mathcal{\bar T}\Big\{\int_0^t d\hat t H^G_{I_{sc}}(\hat t)J^{\nu}_{SCET,I_{sc}}(x_2)\Big\}e^{iH^{SCET}t}
\eea

By following the same series of steps as described in \cite{Vaidya:2020lih}, we can write the result in terms of the free theory interaction picture.
\bea
\sigma_{R}^{(2)}(t) &=&|C(Q)|^2I^{\mu \nu}\int d\tilde x e^{-iH^{0}t}\mathcal{T}\Big\{e^{-i\int_0^{t}dt' H_{\text{int},I}(t')}\int_0^t dt_a H^G_{I}(t_a)J^{\mu}_{SCET,I}(x_1)\Big\}|0\rangle \nn\\
&\times& \langle 0|\mathcal{\bar T}\Big\{e^{-i\int_0^{t}dt' H_{\text{int},I}(t')}\int_0^t dt_b H^G_{I}(t_b)J^{\nu}_{SCET,I}(x_2)\Big\}e^{iH^0t}\nn
\eea
where we have written
\bea
 H^{SCET} =H^0+H_{\text{int}}, \ \ \ \ \text{and} \ \ \   \mathcal{O}_I(t) = e^{iH^0t}\mathcal{O}(t)e^{-iH^0t} 
\eea
The next step is to obtain a factorized formula in terms of our EFT modes. 

We can now take the trace over the density matrix inserting our measurement as before. 
\bea
&&\langle X |\mathcal{M} \sigma_R^{(2)}(t\rightarrow \infty)|X \rangle \equiv \Sigma^{(2)}_R \nn\\
&=& |C(Q)|^2I_{\mu \nu}\int d\tilde x
\langle X|\mathcal{T}\Big\{e^{-i\int_0^{t}dt' H_{int,I}(t')}\int_0^t dt_a H^G_{I}(t_a)J^{\mu}_{SCET,I}(x_1)\Big\}|0\rangle \nn\\
&\times& \langle 0|\mathcal{\bar T}\Big\{e^{-i\int_0^{t}dt' H_{int,I}(t')}\int_0^t dt_b H^G_{I}(t_b)J^{\nu}_{SCET,I}(x_2)\Big\}\mathcal{M}|X\rangle \nn
\eea

We want to examine the scenario where the jet interacts with the medium for a finite time $\{0,t_M\}$. In particular, we want to examine the case when this time scale is comparable to or smaller than the formation time of the jet, $t_F$.  In terms of our kinematic scales this corresponds to $t_M  \leq  \lambda^2/Q$, where $\lambda \sim q_T/Q$ is the expansion parameter of our EFT. 

To do this, we explicitly put in the form of our Glauber operator,  considering the case of collinear partons interacting with the soft partons in the medium. These operators were defined in Eq. \ref{EFTOp}.
\bea
H^{G}(t) =\sum_{i,j \in {q,g}} C_{ij} \int d^3 \vec{x} O_{n,i}^A(\vec{x},t) \frac{1}{\mathcal{P}_{\perp}^2}O_{S,j}^A(\vec{x},t) \Theta(t_M-t)
\eea

We have put in the $\Theta$ function in time to turn off the jet-medium interaction at time $t_M$.
As per our assumption,  we can factorize the vacuum soft function from the medium soft function which is defined as 
\bea
S_G^{AB}(x,y) =  \text{Tr}\Big[\frac{1}{\mathcal{P}_{\perp}^2}O_{S}^A(x)\rho_B \frac{1}{\mathcal{P}_{\perp}^2}O_{S}^A(y)\Big]
\label{eq:Sft}
\eea
where the trace is over all final soft states and color.
 The Wilson coefficient $C_G$ for all the Glauber operators is identical,
\bea
C_G =  8\pi \alpha_s
\eea
and allows us to define
\bea
O_n^A = \sum_i O^A_{n,i} , \ \ O_S^B = \sum_j O^{B}_{S,j}
\eea
Since the Glauber exchange in instantaneous in time and probes the medium over a length scale $ \sim 1/k_{\perp} \sim 1/T \ll t_M$,  the soft modes effectively see an infinite medium.  If we assume the medium is homogeneous over its lifetime, then we can use translational invariance to write the medium soft function as 
\bea
 S_G^{AB}(x,y) = \int \frac{d^4 k}{k_{\perp}^4} e^{ik \cdot( x- y)} D_>^{AB}(k)  
 \label{eq:soft}
\eea
The function $D_>^{AB}(k) $ is independent of $k^+$.  The condition of homogeneity is an idealization which is not strictly necessary for proving factorization.  I will comment on how inhomogeneity can be incorporated in this formalism in Section \ref{sec:Conclusion}.

We can now follow the same series of steps as for the leading order term, and apply power counting to measurement functions as well as the momentum conserving $\delta$ functions based on the momentum scaling of each mode.  Accordingly, we factorize the Hilbert space of the final states in terms of the momentum scaling of the modes and pull out the co-ordinate dependence of each mode by acting with the operators on the final state. This yields the following co-ordinate integrals
\bea
I&=&\int dk^+ \int d^4x_1 e^{-ix_1\cdot(p_e+p_{\bar e}-p_{Jn,1}-p_{JS,1}-p_{J\bar n})}\int d^4x_2 e^{-ix_2\cdot(p_e+p_{\bar e}-p_{Jn,2}-p_{JS,2}-p_{J\bar n})} \nn\\
&&\int_0^{t_M}dx_a^0\int d^3x_a e^{-ix_a \cdot (p_{Gn,1})}\times \int_0^{t_M}dx_b^0\int d^3x_b e^{-ix_b \cdot (p_{Gn,2})}e^{ik \cdot( x_a- x_b)}
\eea
where the subscripts G, J tell us whether the momentum is coming from the action of the Glauber Hamiltonian fields or the SCET current respectively.   
The integrals over $x_1$, $x_2$ can be done directly taking the limit $t \rightarrow \infty$ as also the integrals over $\vec{x}_a^{\perp}$, $\vec{x}_b^{\perp}$.   

Given $x_{a/b}^0 \in \{0,t_M\}$ ,$x_{a/b}^z \in \{-\infty, \infty\}$,  we use
\bea 
x^+=\frac{x^0-x^z}{2} , \ \ \ x^- = \frac{x^0+x^z}{2}
\eea
which then translates to the limits 
\bea
x^+ \in \{-\infty, \infty \}, \ \ \ \ x^- \in \{-x^+, -x^++t_M \} 
\eea
 Looking at the integrals over $k^+$, $x_a$ , $x_b$
\bea
 \bar I&=&\int dk^+\int_0^{t_M}dx_a^0 \int d^3x_a \int_0^{t_M}dx_b^0 \int d^3x_b e^{i k\cdot (\hat x_a- \hat x_b)}e^{i x\cdot p_{GnA}}e^{-ix_b\cdot p_{GnB}}\nn\\
&=& \int dk^+\int_{-\infty}^{\infty}dx_a^+ \int_{-x_a^+}^{-x_a^++t_M} dx_a^- \int_{-\infty}^{\infty}dx_b^+ \int_{-x_b^+}^{-x_b^++t_M} dx_b^- \Big[e^{i k^+(x_a^-- x_b^-)}\Big]\Big[e^{i x_a^+ p^-_{GnA}+ix_a^-p^+_{GnA}}\Big]\nn\\
&&\Big[e^{-ix_b^+ p^-_{GnB}-ix_b^-p^+_{GnB}}\Big]\delta^2(\vec{k}^{\perp}+\vec{p}^{\perp}_{GnA})\delta^2(\vec{k}^{\perp}+\vec{p}^{\perp}_{GnB})
\eea
We can now do the integrals over $x_a^-$, $x_b^-$, which gives us 
\bea
\bar I &=&  \int dk^+\int_{-\infty}^{\infty}dx_a^+ \int_{-\infty}^{\infty}dx_b^+ \Bigg[e^{-ix_a^+(k^++p_{GnA}^+)}\frac{e^{it_M(k^++p_{GnA}^+)}-1}{i(k^++p_{GnA}^+)}\Bigg]\Bigg[e^{ix_b^+(k^++p_{GnB}^+)}\frac{e^{-it_M(k^++p_{GnB}^+)}-1}{-i(k^++p_{GnB}^+)}\Bigg]\nn\\
&&\Big[e^{i x_a^+ p^-_{GnA}}\Big]\Big[e^{-ix_b^+ p^-_{GnB}}\Big]\delta^2(\vec{k}^{\perp}+\vec{p}^{\perp}_{GnA})\delta^2(\vec{k}^{\perp}+\vec{p}^{\perp}_{GnB})
\eea
Next we want to do the integrals over $x_a^+$,$x_b^+$, applying power counting for Glauber momentum  which allows us to drop the terms $e^{-ix_a^+(k^++p_{GnA}^+)}$  and $e^{ix_b^+(k^++p_{GnB}^+)}$ we write 
\small
\bea
 \bar I &=&  \int dk^+\Bigg[\frac{e^{it_M(k^++p_{GnA}^+)}-1}{i(k^++p_{GnA}^+)}\Bigg]\Bigg[\frac{e^{-it_M(k^++p_{GnB}^+)}-1}{-i(k^++p_{GnB}^+)}\Bigg]\delta( p^-_{GnA})\delta(p^-_{GnB})\delta^2(\vec{k}^{\perp}+\vec{p}^{\perp}_{GnA})\delta^2(\vec{k}^{\perp}+\vec{p}^{\perp}_{GnB})\nn\\
&=&  4\pi\left(\frac{t_M}{2}\right)\text{sinc}[(p_{GnA}^+-p_{GnB}^+)\frac{t_M}{2}]e^{-i(p_{GnB}^+-p_{GnA}^+)\frac{t_M}{2}}\delta( p^-_{GnA})\delta(p^-_{GnB})\delta^2(\vec{k}^{\perp}+\vec{p}^{\perp}_{GnA})\delta^2(\vec{k}^{\perp}+\vec{p}^{\perp}_{GnB})\nn
\eea
\normalsize
Now performing the integrals and applying power counting , we have 
\small
\bea
I&=&\delta(Q-p^-_{Jn1})\delta(Q-p_{\bar n}^+)\delta^2(p^{\perp}_{Jn,1}+p^{\perp}_{Js,1}+p^{\perp}_{\bar n}) \delta(Q-p^-_{Jn2})\delta(Q-p_{\bar n}^+)\delta^2(p^{\perp}_{Jn,2}+p^{\perp}_{Js,2}+p^{\perp}_{\bar n})\nn\\
&& 4\pi\left(\frac{t_M}{2}\right)\text{sinc}[(p_{GnA}^+-p_{GnB}^+)\frac{t_M}{2}]e^{-i(p_{GnB}^+-p_{GnA}^+)\frac{t_M}{2}}\delta( p^-_{GnA})\delta(p^-_{GnB})\delta^2(\vec{k}^{\perp}+\vec{p}^{\perp}_{GnA})\delta^2(\vec{k}^{\perp}+\vec{p}^{\perp}_{GnB})\nn
\eea
\normalsize
This simplification follows from the idea that $p_{GS}$ scales as the Glauber momentum. We have ignored any factors of $2\pi$ which will be absorbed in the overall co-efficient for $\Sigma_R^{(2)}$.  The analysis above is physically intuitive as it suggests that the "+" component of the momentum at the Glauber vertex is not conserved if $t_M \leq 1/p^+\sim Q\lambda^2$, which is the scenario we want to consider.  

We also have additional constraints since the total momentum for a particular mode must match on both sides of the cut so that
\bea
p^-_{Jn,1}&=& p^-_{Jn,2}, \ \ \ \ p^{\perp}_{Jn,1}= p^{\perp}_{Jn,2},  \  \ \text{from which it follows that} \ \ \ \ p_{JS,1}= p_{JS,2}\nn
\eea
We can simplify our measurement $\delta$ functions using these set of constraints 
\bea
&&\delta^2(\vec{q}_T- p^{\perp}_{n,\in \text{gr}}-p^{\perp}_{\bar n, \in \text{gr}})\equiv \delta^2(\vec{q}_T- p^{\perp}_{n,\in \text{gr}}-p^{\perp}_{\bar n, \in \text{gr}})\nn\\
&=&  \delta^2(\vec{q}_T- p^{\perp}_{Gn,1 \in \text{gr}}- p^{\perp}_{Jn,1 \in \text{gr}}-p^{\perp}_{\bar n,\in \text{gr}})= \delta^2(\vec{q}_T +\vec{k}_{\perp}+p^{\perp}_{JS}+p^{\perp}_{n,\not\in \text{gr}}+p^{\perp}_{\bar n,\not\in \text{gr}})
\eea

As for the leading order term, we can set the axis of the $\bar n$ jet  to be exactly aligned with the $\bar n$ axis  and then using RPI I, do the same for the n jet before it interacts with the medium.  Using the rest of the constraints then, once again we have an overall factor of $V$. 
\bea 
\label{SigmA}
\Sigma^{(2)}_R(\vec{q}_T,e_n,e{\bar n}) &=& V\times |C_G|^2 H(Q,\mu) S(\vec{q}_T)\otimes_{q_T}\mathcal{J}^{\perp}_{\bar n}(e_{\bar n}, \vec{q}_T)\nn\\
&& \otimes_{q_T}\int \frac{d^2 k_{\perp}dk^-}{k_{\perp}^4}D_>^{AB}(k_{\perp},k^-) \mathcal{J}_n^{AB}(e_n, \vec{q}_T, \vec{k}_{\perp})  
\eea

The convolution in $\vec{q}_T $ turns into a product in impact parameter space. We can now divide and multiply by the vacuum jet function $\mathcal{J}^{\perp}_n(e_n, \vec{b})$ which allows us to factor out the vacuum cross section and define a medium structure function and a medium induced jet function
\bea
 \Sigma^{(2)}_R(\vec{q}_T,e_n,e_{\bar n})=  |C_G|^2 \int d^2\vec{b} e^{i \vec{b} \cdot \vec{q}_T}\Sigma^{(0)}(b,e_n,e_{\bar n})\int d^2 k_{\perp}S_G^{AB}(k_{\perp}) J_{n,R}^{AB}(e_n, \vec{b}, \vec{k}_{\perp})
\eea
where 
\bea
  J_{n,R}^{AB}(e_n, \vec{b}, \vec{k}_{\perp}) = \frac{1}{k_{\perp}^2}\frac{\mathcal{J}_n^{AB}(e_n, \vec{b}, \vec{k}_{\perp})}{\mathcal{J}^{\perp}_n(e_n, \vec{b})}\nn\\
	S_G^{AB}(k_{\perp}) = \int dk^-\frac{1}{k_{\perp}^2}D_>^{AB}(k_{\perp},k^-)
	\label{Fact_Def}
\eea
where the soft function is defined in Eq. \ref{eq:soft}.
\bea
 \mathcal{J}_n^{AB}(\vec{q}_T, \vec{k}_{\perp}) &=&t_M\int dp_{GnA}^+\int dp_{GnB}^+\text{sinc}[(p_{GnA}^+-p_{GnB}^+)\frac{t_M}{2}]e^{-i(p_{GnB}^+-p_{GnA}^+)\frac{t_M}{2}} \nn\\
 &&\langle X_n|\mathcal{T}\Big\{[\delta^2(\vec{k}_{\perp}+\mathcal{P}_{\perp})\delta(\mathcal{P}^-)\delta(p_{GnA}^+-\mathcal{P}^+)O_n^A(0)]\bar \chi_n(0)\Big\}\frac{\slashed{\bar{n}}}{2}|0\rangle\nn\\
&&\langle 0|\bar{\mathcal{T}}\Big\{[\delta^2(\mathcal{P}_{\perp})\chi_n(0)][\delta^2(\vec{k}_{\perp}+\mathcal{P}_{\perp})\delta(\mathcal{P}^-)\delta(p_{GnB}^+-\mathcal{P}^+)O_n^B(0)]\delta^2(Q-\mathcal{P}^-)\mathcal{M}|X_n\rangle \nn\\
\label{eq:jetr}
\eea
$\mathcal{P}^{\mu}$ is a momentum operator that extracts out the $\mu$ component of momentum from the operator it acts on. $\mathcal{M}$ is the measurement function acting on the final state $X_n$
\bea
\mathcal{M} = \delta^2(\vec{q}_{T} - \sum_i \vec{p}^X_{i\not \in \text{gr}}-\vec{k}_{\perp})\Theta\left(e_n- \frac{4}{Q^2}\left(\sum_{i \in \text{gr}}p^X_i \right)^2\right)
\eea

\subsubsection{Glauber insertion on same side of the cut}

We now consider the structure for inserting the Glauber interaction on the same side of the cut.  Starting with our master formula in Eq.\ref{GExpp} and making two Glauber insertions on the same side of the cut, we have
\bea
\Sigma_V^{(2)}&=&  \frac{(-i)^2}{2!}\int d^4y_1 \int d^4y_2  |C(Q)|^2 I_{\mu\nu}\int d \tilde x e^{-iH^{SCET}t}\mathcal{T} \Big\{  H^G_{I_{sc}}(y_1) H^G_{I_{sc}}(y_2)J^{\mu}_{SCET,I_{sc}}(x_1)\Big\} |0\rangle \nn\\
&& \langle 0| \bar{\mathcal{T}} \Big\{J^{\nu}_{SCET,I_{sc}}(x_2) \Big\} e^{iH^{SCET}t}\nn\\
&+&  \frac{(-i)^2}{2!}\int d^4y_1 \int d^4y_2  |C(Q)|^2 I_{\mu\nu}\int d \tilde x e^{-iH^{SCET}t}\mathcal{T} \Big\{ J^{\mu}_{SCET,I_{sc}}(x_1)\Big\} |0\rangle \nn\\
&& \langle 0| \bar{\mathcal{T}} \Big\{ H^G_{I_{sc}}(y_1) H^G_{I_{sc}}(y_2)J^{\nu}_{SCET,I_{sc}}(x_2) \Big\} e^{iH^{SCET}t}
\eea
 
We can now write this in the free theory interaction picture 
\bea
\Sigma_V^{(2)}&=&  \frac{(-i)^2}{2!}\int d^4y_1 \int d^4y_2  |C(Q)|^2 I_{\mu\nu}\int d \tilde x e^{-iH^{0}t} \mathcal{T} \Big\{ e^{-i\int_0^t dt'H_{\text{int},I}(t')} H^G_{I}(y_1) H^G_{I}(y_2)J^{\mu}_{SCET,I}(x_1)\Big\} |0\rangle \nn\\
&& \langle 0| \bar{\mathcal{T}} \Big\{e^{-i\int_0^t dt'H_{\text{int},I}(t')}J^{\nu}_{SCET,I}(x_2) \Big\} \nn\\
&+&  \frac{(-i)^2}{2!}\int d^4y_1 \int d^4y_2  |C(Q)|^2 I_{\mu\nu}\int d \tilde x \mathcal{T} \Big\{e^{-i\int_0^t dt'H_{\text{int},I}(t')} J^{\mu}_{SCET,I}(x_1)\Big\} |0\rangle \nn\\
&& \langle 0| \bar{\mathcal{T}} \Big\{ e^{-i\int_0^t dt'H_{\text{int},I}(t')}H^G_{I}(y_1) H^G_{I}(y_2)J^{\nu}_{SCET,I}(x_2) \Big\} e^{iH^{0}t} 
\eea
Next we separate out the Soft fields from the collinear further factorizing the vacuum soft function from the medium soft function.
We also take a trace over the final states while putting in the measurement. 
\bea
\Sigma_V^{(2)}&=&  \frac{(-i)^2}{2!}\int d^4y_1 \int d^4y_2  |C(Q)|^2 I^{\mu\nu}\int d \tilde x S(x_1,  x_2)J_{\bar n}(x_1, x_2) S^{AB}_{G,V}(y_1,  y_2) J^{AB}_{n,V}(x_1, x_2, y_1, y_2) \nn\\
&+&  \frac{(-i)^2}{2!}\int d^4y_1 \int d^4y_2  |C(Q)|^2 I^{\mu\nu}\int d \tilde x S(x_1,  x_2)J_{\bar n}(x_1, x_2) \bar{S}^{AB}_{G,V}(y_1,  y_2) \bar{J}^{AB}_{n,V}(x_1, x_2, y_1, y_2)\nn
\eea

with the definitions 
\bea
 S_{G,V}^{AB}(y_1,y_2) = \langle X_S| \mathcal{T} \Big\{ e^{-i\int_0^t dt'H_{S\text{int},I}(t')}O_{S,I}^A(y_1)O_{S,I}^B(y_2)\Big\}\rho_B\bar{\mathcal{T}}\Big\{ e^{-i\int_0^t dt'H_{S\text{int},I}(t')}\Big\}|X_s\rangle
\eea
where $H_{Sint}$ is the Soft interaction Hamiltonian.  
\bea
  J^{AB}_{n,V}(x_1, x_2, y_1, y_2)&=& \langle X_n| \mathcal{T} \Big\{ e^{-i\int_0^t dt'H_{n\text{int},I}(t')}O_{n,I}^A(y_1)O_{n,I}^B(y_2)\bar {\chi}_n(x_1)\Big\}|0\rangle \nn\\
  &&\langle 0|\bar{\mathcal{T}}\Big\{ e^{-i\int_0^t dt'H_{n\text{int},I}(t')}\frac{\slashed{\bar n}}{2}\chi_n(x_2)\Big\}\mathcal{M}|X_n\rangle\nn
\eea

and similarly $\bar J^{AB}_{n,V}$.
We can do a series of manipulations to express the soft function in terms of the medium structure function $S_G^{AB}$ defined in Eq. \ref{eq:soft}.  We will use the fact the  no measurement is directly imposed on the medium Soft function $S_{G,V}^{AB}$, 
\bea
 S_{G,V}^{AB}(y_1,y_2)&=& \text{Tr}\Big[\mathcal{T}\Big\{e^{-i\int_{y_1^0}^t dt'H_{S\text{int},I}(t')}\Big\}O_{S,I}^A(y_1)\mathcal{T}\Big\{e^{-i\int_{y_2^0}^{y_1^0} dt'H_{S\text{int},I}(t')}\Big\}O_{S,I}^B(y_2)\mathcal{T}\Big\{e^{-i\int_0^{y_2^0} dt'H_{S\text{int},I}(t')}\Big\}\rho_B\nn\\
 &&\bar{\mathcal{T}}\Big\{ e^{-i\int_0^t dt'H_{S\text{int},I}(t')}\Big\}\Big] \Theta(y_1^0-y_2^0)\nn\\
 &+& \text{Tr}\Big[\mathcal{T}\Big\{e^{-i\int_{y_2^0}^t dt'H_{S\text{int},I}(t')}\Big\}O_{S,I}^B(y_2)\mathcal{T}\Big\{e^{-i\int_{y_1^0}^{y_2^0} dt'H_{S\text{int},I}(t')}\Big\}O_{S,I}^A(y_1)\mathcal{T}\Big\{e^{-i\int_0^{y_1^0} dt'H_{S\text{int},I}(t')}\Big\}\rho_B\nn\\
 &&\bar{\mathcal{T}}\Big\{ e^{-i\int_0^t dt'H_{S\text{int},I}(t')}\Big\}\Big] \Theta(y_2^0-y_1^0)
\eea
which can be written as 
\bea
  S_{G,V}^{AB}(y_1,y_2)&=& \text{Tr}\Big[\mathcal{T} \Big\{ e^{-i\int_0^t dt'H_{S\text{int},I}(t')}O_{S,I}^B(y_2)\Big\}\rho_B\bar{\mathcal{T}}\Big\{ e^{-i\int_0^t dt'H_{S\text{int},I}(t')}O_{S,I}^A(y_1)\Big\}\Big]\Theta(y_1^0-y_2^0)\nn\\
 &+& \text{Tr}\Big[\mathcal{T} \Big\{ e^{-i\int_0^t dt'H_{S\text{int},I}(t')}O_{S,I}^A(y_1)\Big\}\rho_B\bar{\mathcal{T}}\Big\{ e^{-i\int_0^t dt'H_{S\text{int},I}(t')}O_{S,I}^B(y_2)\Big\}\Big]\Theta(y_2^0-y_1^0)\nn\\
 &\equiv &   \Theta(y_1^0-y_2^0)S_G^{BA}(y_2,y_1)+  \Theta(y_2^0-y_1^0)S_G^{AB}(y_1,y_2)
\eea

where $S_G^{AB}(x,y)$ is defined in Eq. \ref{eq:Sft}.
Therefore, if we look at the combination $S_{G,V}^{AB}J_{n,V}^{AB}$,  
\bea
 &&\int d^4y_1 \int d^4y_2S_{G,V}^{AB}(y_1,y_2)J_{n,V}^{AB}(x_1,x_2,y_1,y_2)\nn\\
 	&=& \int d^4y_1 \int d^4y_2\Big[ \Theta(y_1^0-y_2^0)S_G^{BA}(y_2,y_1)+  \Theta(y_2^0-y_1^0)S_G^{AB}(y_1,y_2)\Big]J_{n,V}^{AB}(x_1,x_2,y_1,y_2)
\eea
Interchanging $y_1 \leftrightarrow y_2$ and $A \leftrightarrow B$ in the first term, we arrive at 
\bea
 &&\int d^4y_1 \int d^4y_2S_{G,V}^{AB}(y_1,y_2)J_{n,V}^{AB}(x_1,x_2,y_1,y_2)\nn\\
 	&=& \int d^4y_1 \int d^4y_2 \Theta(y_2^0-y_1^0)S_G^{AB}(y_1,y_2)\Big[ J_{n,V}^{AB}(x_1,x_2,y_1,y_2)+J_{n,V}^{BA}(x_1,x_2,y_2,y_1)\Big]\nn\\
 	&=& 2\int d^4y_1 \int d^4y_2 \Theta(y_2^0-y_1^0)S_G^{AB}(y_1,y_2) J_{n,V}^{AB}(x_1, x_2, y_1,  y_2)
 	\eea
 	where we use the fact that $S_G^{AB} \propto \delta^{AB}$.
 	Applying the same logic for the term on the other side of the cut, we can write 
 	\bea
 	 \Sigma_V^{(2)}&=&  -\int d^4y_1 \int d^4y_2  |C(Q)|^2 I^{\mu\nu}\int d \tilde x S(x_1,  x_2)J_{\bar n}(x_1, x_2) S^{AB}_{G}(y_1,  y_2)\nn\\
 	 &&\Big[\Theta(y_2^0-y_1^0) J^{AB}_{n,V}(x_1, x_2, y_1, y_2)+\Theta(y_1^0-y_2^0)  \bar{J}^{AB}_{n,V}(x_1, x_2, y_1, y_2)\Big]\nn
 	\eea
 		We can first write 
 	\bea
 	  S^{AB}_{G}(y_1,  y_2) = \int \frac{d^2k_{\perp} }{k_{\perp}^4}D_>^{AB}(k)\int dk^+ e^{ik \cdot(\hat{y}_1-\hat{y}_2)}
 	\eea
We can do the co-ordinate integrals over $x_1, x_2$ in the same way as for the previous case which then leads to  	
 	\bea
 	  \Sigma_V^{(2)}&=&  -V\times |C_G|^2H(Q,\mu)S(\vec{q}_T)\otimes_{q_T} \mathcal{J}_{\bar n}^{\perp}(e_{\bar n},\vec{q}_T)\nn\\
 	  &\otimes_{q_T}& \int \frac{d^2k_{\perp} }{k_{\perp}^4}D_>^{AB}(k)\mathcal{J}_{n,V}^{AB}(e_n, \vec{q}_T, \vec{k}_{\perp})
 	\eea
 	
 	Lets now look at the "virtual" jet function 
 	\small
 	\bea
 	\mathcal{J}^{AB}_{n,V}&=& \int dk^+ \int d^4y_1 \int d^4y_2 e^{ik \cdot(\hat{y}_1-\hat{y}_2)} \Theta(y_2^0-y_1^0) J^{AB}_{n,V}(\vec{q}_T,e_n,  y_1,  y_2)+\Theta(y_1^0-y_2^0)  \bar{J}^{AB}_{n,V}(x_1, x_2, y_1, y_2)\nn
 	\eea
 	\normalsize
 We can write the $\Theta$ function  in the integral representation 
\bea
 \Theta(y_i^0-y_j^0)=\frac{1}{2\pi i} \int_{-\infty}^{\infty} \frac{dw e^{iw (y_i^0-y_j^0)}}{w-i\epsilon}
\eea
which then allows us to perform the same series of steps as for the Glauber insertion on opposite sides of the cut, 
\small
\bea
&&\mathcal{J}_{n,V}^{AB}(\vec{q}_T, \vec{k}_{\perp}) = t_M\frac{1}{2\pi i} \int_{-\infty}^{\infty} \frac{dw}{w-i\epsilon}\Bigg\{
\int dp_{GnA}^+\int dp_{nB}^+ \text{sinc}[(p_{GnA}^++p_{GnB}^+)\frac{t_M}{2}]e^{-i(p_{GnB}^++p_{GnA}^+)\frac{t_M}{2}}\nn\\
&&\langle X_n|\mathcal{T}\Big\{[\delta^2(\vec{k}_{\perp}+\mathcal{P}_{\perp})\delta(\mathcal{P}^-)\delta(p_{GnA}^+-\mathcal{P}^+)O_n^A(0)][\delta^2(\vec{k}_{\perp}+\mathcal{P}_{\perp})\delta(\mathcal{P}^-)\delta(p_{GnB}^+-\mathcal{P}^+)O_n^B(0)]\bar \chi_n(0)\Big\}\frac{\slashed{\bar{n}}}{2}|0\rangle\nn\\
&&\langle 0|\bar{\mathcal{T}}\Big\{[\delta^2(\mathcal{P}_{\perp})\chi_n(0)]\delta^2(Q-\mathcal{P}^-)\Big\}\mathcal{M}|X_n\rangle +c.c \Bigg\}\
 	\eea
 \normalsize
We can now do the integral over w which simply give us a factor of (1/2) for each term so that our result becomes 
\small
\bea 
&&\mathcal{J}_{n,V}^{AB}(\vec{q}_T, \vec{k}_{\perp}) = t_M\frac{1}{2}\Bigg\{
\int dp_{GnA}^+\int dp_{nB}^+ \text{sinc}[(p_{GnA}^++p_{GnB}^+)\frac{t_M}{2}]e^{-i(p_{GnB}^++p_{GnA}^+)\frac{t_M}{2}}\nn\\
&&\langle X_n|\mathcal{T}\Big\{[\delta^2(\vec{k}_{\perp}+\mathcal{P}_{\perp})\delta(\mathcal{P}^-)\delta(p_{GnA}^+-\mathcal{P}^+)O_n^A(0)][\delta^2(\vec{k}_{\perp}+\mathcal{P}_{\perp})\delta(\mathcal{P}^-)\delta(p_{GnB}^+-\mathcal{P}^+)O_n^B(0)]\bar \chi_n(0)\Big\}\frac{\slashed{\bar{n}}}{2}|0\rangle\nn\\
&&\langle 0|\bar{\mathcal{T}}\Big\{[\delta^2(\mathcal{P}_{\perp})\chi_n(0)]\delta^2(Q-\mathcal{P}^-)\Big\}\mathcal{M}|X_n\rangle +c.c \Bigg\}
\label{eq:jetv}
\eea 	
\normalsize

We can now write the $\Sigma_V^{(2)}$  in a factorized form as 
\bea
 \Sigma^{(2)}_V(\vec{q}_T,e_n,e_{\bar n})=  -|C_G|^2 \int d^2\vec{b} e^{i \vec{b} \cdot \vec{q}_T}\Sigma^{(0)}(b,e_n,e_{\bar n})\int d^2 k_{\perp}S_G^{AB}(k_{\perp}) J_{n,V}^{AB}(e_n, \vec{b}, \vec{k}_{\perp})
\eea
where 
\bea
  J_{n,V}^{AB}(e_n, \vec{b}, \vec{k}_{\perp}) = \frac{1}{k_{\perp}^2}\frac{\mathcal{J}_{n,V}^{AB}(e_n, \vec{b}, \vec{k}_{\perp})}{\mathcal{J}^{\perp}_n(e_n, \vec{b})}\nn\\
	S_G^{AB}(k_{\perp}) = \int dk^-\frac{1}{k_{\perp}^2}D^{AB}(k_{\perp},k^-)
	\label{Fact_Def}
\eea

We can now combine the  all the pieces upto quadratic order in Glauber insertion to write a very simple final factorized form for the density matrix as 
\begin{tcolorbox}
\small
\bea
 \Sigma(\vec{q}_T,e_n,e_{\bar n})=  \Sigma^{(0)}(\vec{q}_T,e_n,e_{\bar n})+ |C_G|^2 \int d^2\vec{b} e^{i \vec{b} \cdot \vec{q}_T}\Sigma^{(0)}(b,e_n,e_{\bar n})\int d^2 k_{\perp}S_G^{AB}(k_{\perp}) J_{n}^{AB}(e_n, \vec{b}, \vec{k}_{\perp})\nn
\eea
\normalsize
\end{tcolorbox}
where $S_G^{AB}(k_{\perp})$ is defined in Eq.\ref{eq:soft}, while the medium induced jet function is defined as 
\bea
 J_{n}^{AB}(e_n, \vec{b}, \vec{k}_{\perp})= J_{n,R}^{AB}(e_n, \vec{b}, \vec{k}_{\perp})-J_{n,V}^{AB}(e_n, \vec{b}, \vec{k}_{\perp})
 \label{eq:jet}
\eea

in terms of the functions defined in Eq. \ref{eq:jetr} and Eq.\ref{eq:jetv}.
We therefore see that the form of the factorized density matrix is identical to the case of a long lived medium derived in \cite{Vaidya:2020lih}. In fact, every function except for the medium jet function has the same definition.  So in the next section we will quote the radiative corrections for the medium soft function and look at the one loop correction for the medium jet function in Section \ref{sec:Loop}.
\section{Radiative corrections for the Medium structure function}

The medium structure function or the "PDF of the medium" is defined in terms of the correlators of the Soft current in the QGP background 
\bea
S_G^{AB}(k_{\perp})=\frac{1}{k_{\perp}^2}\int \frac{dk^-}{2\pi} D_>^{AB}(k_{\perp},k^-)
\eea
where 
\bea
D_>^{AB}(k) = \int d^4x e^{-ik\cdot x}\langle X_S|O_S^A(x)\rho O_S^B(0)|X_S\rangle  
\eea

The soft current is defined in Eq.\ref{eq:Sft}.

This function is identical to the medium structure function that appears for the case of a dilute long lived medium considered in \cite{Vaidya:2021vxu}.  The tree level and one loop results for this function for a thermal medium were computed in that paper.  Here  I will only quote the Renormalization Group(RG) equations which result from the loop calculation in the background of a thermal density matrix. Note that the RG equations are independent of the form of the medium density matrix. There are two types of divergences that appear at one loop: UV divergences induced by the separation of the Soft scale T from the hard scale Q and rapidity divergences which appear due to the factorization of energetic collinear partons from the Soft partons. This requires us to introduce a renormalization scale for each of these divergences namely $\nu,$ for rapidity and $\mu$ for UV poles. 

The rapidity RG equation for the Soft function at this order in perturbation theory is 
 \bea
\nu\frac{d}{d\nu} S(\vec{k}_{\perp})= \frac{\alpha_sN_c}{\pi^2}\int d^2q_{\perp} \left( \frac{S(\vec{q}_{\perp})}{(\vec{q}_{\perp}-\vec{k}_{\perp})^2} -\frac{k_{\perp}^2S(k_{\perp})}{2q_{\perp}^2(\vec{q}_{\perp}-\vec{k}_{\perp})^2}\right)
\eea
This is just the BFKL equation. The UV RG equation is simply the running of the QCD coupling.
\bea
 \mu\frac{d}{d\mu} S(\vec{k}_{\perp})&=& \frac{\alpha_s\beta_0}{\pi}S(k_{\perp})
\eea
where $\beta_0 = 11/3 C_A - C_Fn_FT_F$, which has a value of 23/3 if we assume 5 active quark flavors.

By RG consistency we require that the medium induced jet function should be UV finite and evolve in rapidity with the negative of the BFKL kernel.  

\section{The medium induced jet function in a short lived medium}
\label{sec:Loop}
The medium induced jet function is defined in Eq. \ref{eq:jet}, 
\bea
 J_{n}^{AB}(e_n, \vec{b}, \vec{k}_{\perp})= J_{n,R}^{AB}(e_n, \vec{b}, \vec{k}_{\perp})-J_{n,V}^{AB}(e_n, \vec{b}, \vec{k}_{\perp})
\eea

The definition of the components $J_{n,V}^{AB}$ and $J_{n,R}^{AB}$  is different from the case of the long lived medium considered in \cite{Vaidya:2020lih,Vaidya:2021vxu} so that we need to recompute this function.  This function knows about the shortened lifetime of the medium $t_M$ which now changes the structure of all the Feynman diagrams computed in \cite{Vaidya:2020lih,Vaidya:2021vxu}.  At the same time it will add new diagrams which incorporate quantum interference between the hard vertex that creates the jet and the subsequent interaction with the medium.  
For the case of the long lived medium, this function was shown to obey a BFKL evolution equation with a natural scale $\nu \sim Q \sim Qz_c$.  Since all other functions in the factorized formula are identical, by RG consistency , we expect the BFKL evolution to hold even in the case of a short lived medium, although the natural scale may now change. 

The great advantage of using an EFT approach now becomes apparent since we only need to recalculate the jet function leaving all the other functions in the factorized formula unchanged compared to the long lived medium. 

We now consider in detail the one loop correction for the jet function.  We consider  the functions $\mathcal{J}^{AB}_{n,R}$ and $\mathcal{J}_{n,V}^{AB}$ in turn. 

\subsection{$J_n^R$}

We give the definition of this function here for convenience
\bea
  J_{n,R}^{AB}(e_n, \vec{b}, \vec{k}_{\perp}) = \frac{1}{k_{\perp}^2}\frac{\mathcal{J}_n^{AB}(e_n, \vec{b}, \vec{k}_{\perp})}{\mathcal{J}^{\perp}_n(e_n, \vec{b})}\nn
  \eea
where  $\mathcal{J}^{\perp}_n(e_n, \vec{b})$ is the vacuum jet function.  At one loop this function was computed in \cite{Vaidya:2020lih}, 
\bea
\mathcal{J}_n^{\perp(1)}(e_n,b,\mu,\nu)&=& -\frac{\alpha_sC_F}{(2\pi)^3} \ln \frac{\mu^2}{E_J^2e_n}\left( 2\ln \frac{1-z_c}{z_c}- 2(1-z_c)+2z_c +\frac{(1-z_c)^2}{2}-\frac{z_c^2}{2}\right) \nn\\
&-&\frac{\alpha_sC_F}{(2\pi)^3}\ln \frac{\mu^2b^2e^{2\gamma_E}}{4} \left(-2\ln \frac{\nu}{Qz_c}-4z_{c}-2\ln (1-z_c)+\frac{z_c^2}{2}+\frac{1}{2}-\frac{(1-z_c)^2}{2} \right)\nn\\
\label{eq:jetvac}
\eea
while $\mathcal{J}_n^{AB}(\vec{q}_T, \vec{k}_{\perp},e_n) $ is defined in Eq.\ref{eq:jetr}.

\subsubsection{Tree level result} 
\begin{figure}
\centering
\includegraphics[width=0.4\linewidth]{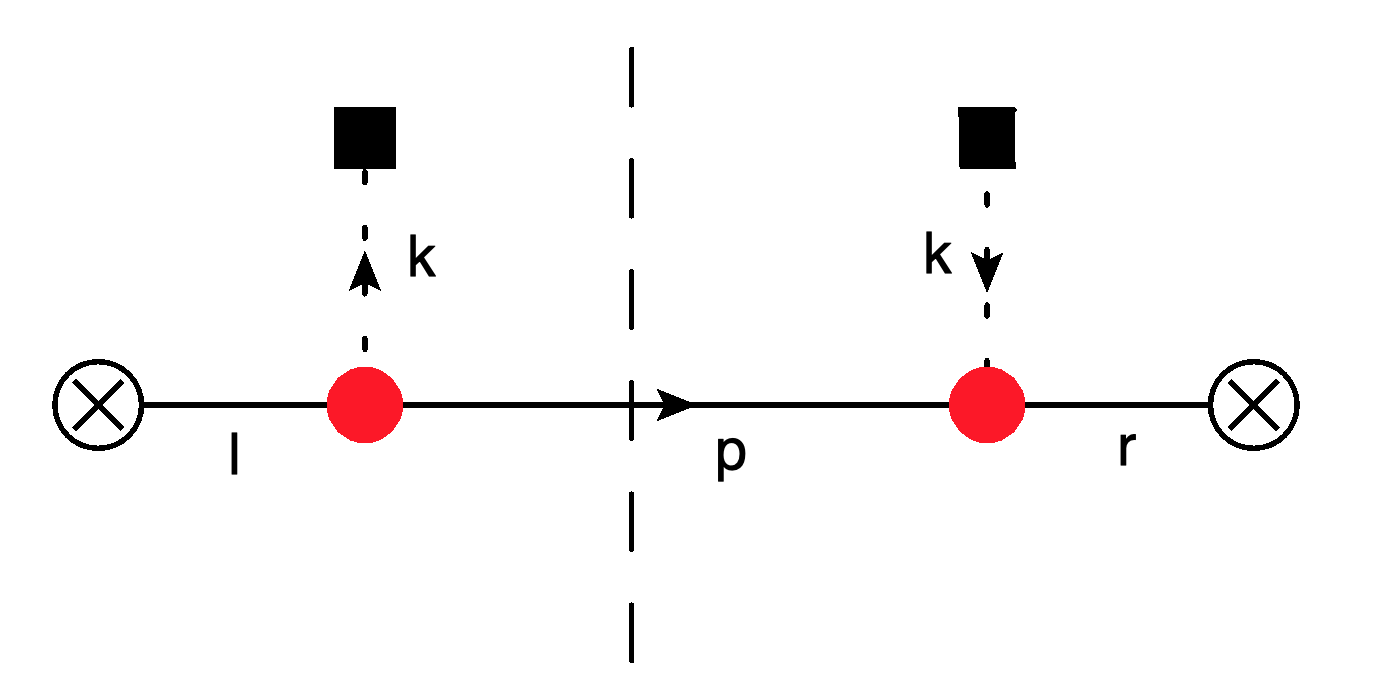}
\caption{Tree level medium jet function. The $\otimes$ vertex indicates the hard vertex.  The red dot is the collinear Glauber vertex while the black square is soft(medium) Glauber vertex.  The dotted line is the Glauber gluon with momentum k.}
\label{Rtree}
\end{figure}

Now lets compute the tree level result.  At tree level we have a single quark as shown in Fig.\ref{Rtree} and we expect in the absence of radiation that the quark is created on-shell. 

\bea
 \mathcal{J}_n^{AB(0)} &=&2 t_M\frac{\delta^{AB}}{2} \int d^4pp^-\delta^+(p^2)\int \frac{d^4l}{2\pi}\frac{(i)l^-}{l^2+i\epsilon}\int \frac{d^4r}{2\pi}\frac{(-i)r^-}{r^2-i\epsilon}\text{sinc}[(l^+-r^+)\frac{t_M}{2}]e^{i(r^+-l^+)\frac{t_M}{2}}\nn\\
&&\delta^2(\vec{k}_{\perp}+\vec{p}_{\perp}-\vec{l}_{\perp})\delta(p^--l^-)\delta^2(\vec{r}_{\perp})][\delta^2(\vec{k}_{\perp}+\vec{p}_{\perp}-\vec{r}_{\perp})\delta(p^--r^-)\delta(Q-r^-)\delta^2(\vec{q}_{Tn}-\vec{k}_{\perp})\nn
\eea
We can do contour integrals over $l^+, r^+$, while eliminating all other integrals using the $\delta$ functions, which leads us to 
\bea
\mathcal{J}_n^{AB(0)} (\vec{q}_{Tn},\vec{k}_{\perp})= t_M\frac{\delta^{AB}}{2}\delta^2(\vec{q}_{Tn}-\vec{k}_{\perp})
\eea
which in impact parameter space becomes 
\bea
 \mathcal{J}_n^{AB(0)} (\vec{b}, \vec{k}_{\perp}) =  t_M\frac{\delta^{AB}}{2}e^{-i \vec{k}_{\perp} \cdot \vec{b}}
\eea


\subsubsection{One Loop}

The total number of diagrams at one loop is quite large and to keep things simple I will only consider only those diagrams that lead to the dominant logarithms. These logarithms appears in the case when the radiated gluon with momentum q  fails grooming and therefore contributes to the transverse momentum imbalance between the jets.  The dominant contribution appears in the limit $q^- \rightarrow 0$ which leads to a rapidity divergence and ultimately to a BFKL evolution.  

We can organize the corrections in terms of real and virtual gluon contributions.   We will present the final results here while relegating the details to Appendix \ref{app:OneLoop}.
\begin{itemize}
\item{Real Gluon emission $\mathcal{J}_{n,R}^R$}

We first consider the set of diagrams which consist of  real Gluon emission.  Fig.\ref{JR} shows Wilson line contributions from the hard vertex.

\begin{figure}
\centering
\includegraphics[width=0.8\linewidth]{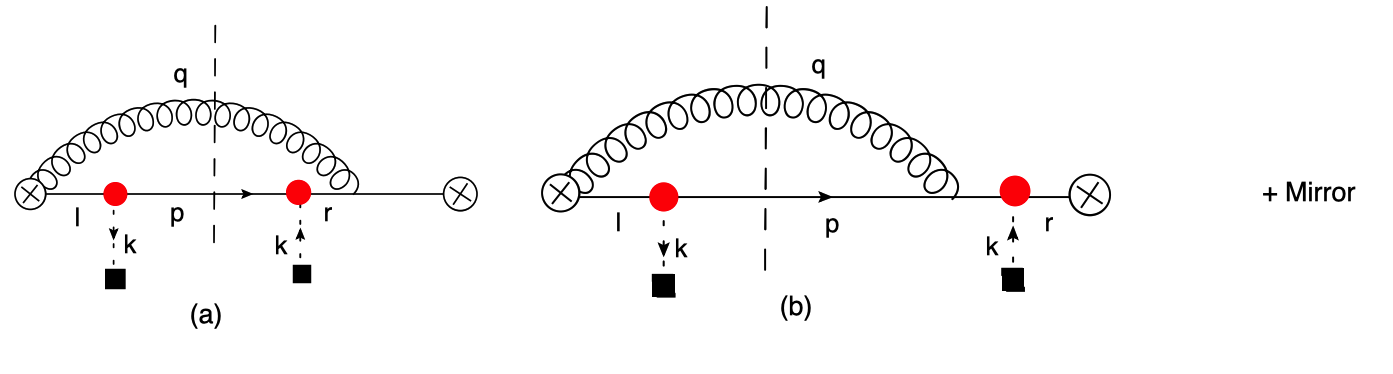}
\caption{The real gluon emission comes from the colinear Wilson line attached to the hard vertex.}
\label{JR}
\end{figure}
We can  add the mirror diagrams and expand around $q^- =0$ ,retaining the possible dominant logarithmic contributions
\bea
2(a)+2(b)&=&   -g^2\delta^{AB}t_M \int \frac{d^{2}q_{\perp}}{(2\pi)^{3}}\frac{\delta^2(\vec{q}_{Tn}-\vec{q}_{\perp}-\vec{k}_{\perp})}{q_{\perp}^2+m_D^2}\Bigg\{C_F\left(\frac{\nu}{Qz_c}\right)^{\eta}\frac{1}{\eta}-\frac{N_C}{2}F\Big[\frac{(q_{\perp}^2+m_D^2)t_M}{Qz_c}\Big]\Bigg\}\nn\\
\eea
where 
\bea
F(x)=\text{CosInt}\Big[x\Big]-\text{Sinc}\Big[x\Big]
\label{eq:F}
\eea
is written in terms of the CosInt function 
\bea
\text{CosInt}[z] = \int_z^{\infty} du \frac{\cos u}{u}
\eea
$\eta$ is the rapidity pole and the corresponding renormalization scale is $\nu$.
Since we are dividing out by the vacuum jet function,  the piece proportional to $C_F$ cancels out  so we will ignore this term in the final result.

We can next consider the remaining diagrams that give a contribution from the hard vertex Wilson line along with the collinear Gluon Glauber vertex. These diagrams are shown in Fig.\ref{djg}. All of them reduce to 0 due to the SCET collinear sector Feynman rules.

\begin{figure}
\centering
\includegraphics[width=\linewidth]{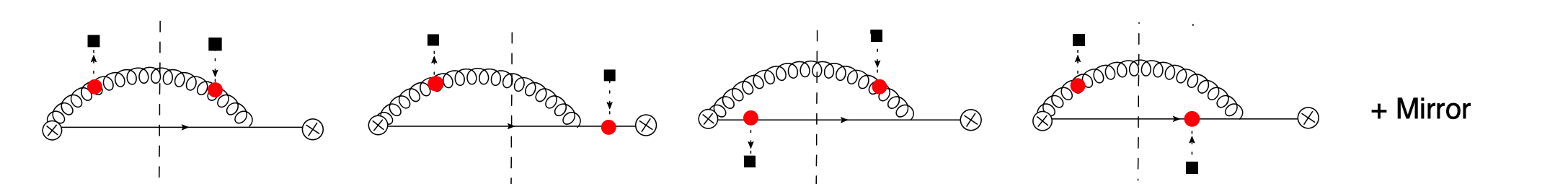}
\caption{Hard vertex Wilson line gluon interacting with medium.}
\label{djg}
\end{figure}

\begin{figure}
\centering
\includegraphics[width=0.9\linewidth]{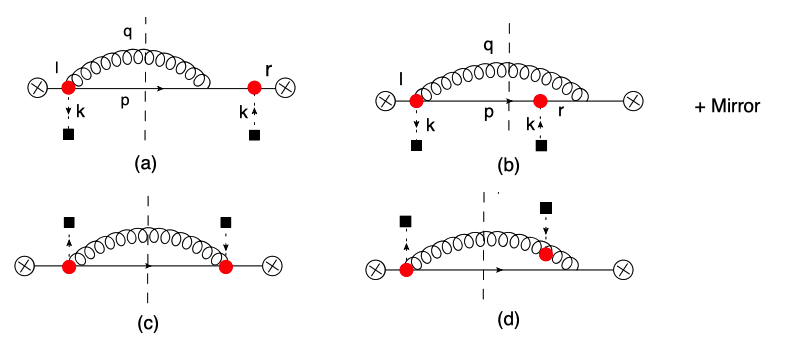}
\caption{Real gluon emission from collinear Wilson line of the Quark Glauber vertex.}
\label{djgw}
\end{figure}
The next set of diagrams are ones which give a Wilson line contribution from the Glauber vertex. There are 4 such diagrams as shown in Fig.\ref{djgw}, two of which ((c) and (d)) reduce to 0.  The diagrams (a) and (b) along with their mirrors add up to  
\small
\bea
 4(a)+4(b)&=& -2g^2t_M\nu^{\eta}N_C\frac{\delta^{AB}}{2}\int \frac{d^2q_{\perp}}{(2\pi)^3}\frac{\delta^2(\vec{q}_{Tn}-\vec{q}_{\perp}-\vec{k}_{\perp})}{q_{\perp}^2+m_D^2}\Bigg\{\left(\frac{\nu}{Qz_c}\right)^{\eta}\frac{1}{\eta}-\frac{1}{2}F\Big[\frac{(q_{\perp}^2+m_D^2)t_M}{Qz_c}\Big]\Bigg\}\nn
\eea 
\normalsize
The final set of diagrams with a rapidity logarithm are those with a collinear Gluon-Glauber interaction along with a Lagrangian vertex insertion (Fig. \ref{djgl}).

\begin{figure}
\centering
\includegraphics[width=0.9\linewidth]{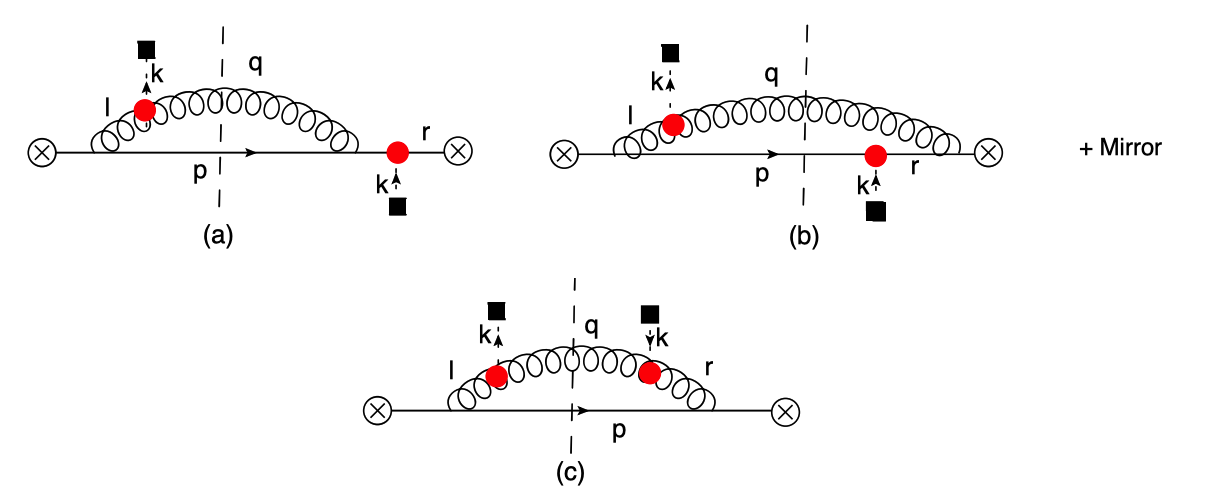}
\caption{Collinear gluon emission with Glauber interaction with the medium.}
\label{djgl}
\end{figure}

The diagrams (a) and (b) add up to 
\bea
 &&5(a)+5(b) = 2g^2N_Ct_M\frac{\delta^{AB}}{2} \int \frac{d^{2}q_{\perp}}{(2\pi)^3}\frac{\delta^2(\vec{q}_{Tn}-\vec{q}_{\perp}-\vec{k}_{\perp})}{q_{\perp}^2+m_D^2}\frac{2\vec{q}_{\perp}\cdot (\vec{q}_{\perp}+\vec{k}_{\perp})}{(\vec{q}_{\perp}+\vec{k}_{\perp})^2} \Bigg\{\left(\frac{\nu}{Qz_c}\right)^{\eta}\frac{1}{\eta}\nn\\
 &-&F\Big[\frac{([\vec{q}_{\perp}+\vec{k}_{\perp}]^2+m_D^2)t_M}{Qz_c}\Big]-\frac{1}{2}F\Big[\frac{(\vec{q}_{\perp}^2+m_D^2)t_M}{Qz_c}\Big]+\frac{1}{2}F\Big[\frac{(\Big[\vec{q}_{\perp}+\vec{k}_{\perp}\Big]^2-q_{\perp}^2)t_M}{Qz_c}\Big]
\Bigg\}\nn
\eea

The final diagram is (c)  which gives us 
\bea
 5(c) &=& -4g^2N_Ct_M\frac{\delta^{AB}}{2} \int \frac{d^{2}q_{\perp}}{(2\pi)^3}\frac{\delta^2(\vec{q}_{Tn}-\vec{q}_{\perp}-\vec{k}_{\perp})}{(\vec{q}_{\perp}+\vec{k}_{\perp})^2} \Bigg\{\left(\frac{\nu}{Qz_c}\right)^{\eta}\frac{1}{\eta}-F\Big[\frac{([\vec{q}_{\perp}+\vec{k}_{\perp}]^2+m_D^2)t_M}{Qz_c}\Big]
\Bigg\}\nn
\eea

All the real gluon emission diagrams for Figs.2 through 5  now sum up to 
\small
\bea
\mathcal{J}_{n,R}^R&=& g^2N_Ct_M\delta^{AB} \int \frac{d^{2}q_{\perp}}{(2\pi)^3}\frac{\delta^2(\vec{q}_{Tn}-\vec{q}_{\perp}-\vec{k}_{\perp})}{q_{\perp}^2+m_D^2}\Bigg[\frac{-k_{\perp}^2}{(\vec{q}_{\perp}+\vec{k}_{\perp})^2} \Bigg\{\left(\frac{\nu}{Qz_c}\right)^{\eta}\frac{1}{\eta}-F\Big[\frac{([\vec{q}_{\perp}+\vec{k}_{\perp}]^2+m_D^2)t_M}{Qz_c}\Big]\Bigg\}\nn\\
&-&F\Big[\frac{([\vec{q}_{\perp}+\vec{k}_{\perp}]^2+m_D^2)t_M}{Qz_c}\Big]+F\Big[\frac{(q_{\perp}^2+m_D^2)t_M}{Qz_c}\Big]\nn\\
 &+&\frac{2\vec{q}_{\perp}\cdot (\vec{q}_{\perp}+\vec{k}_{\perp})}{(\vec{q}_{\perp}+\vec{k}_{\perp})^2} \Bigg\{-\frac{1}{2}F\Big[\frac{(\vec{q}_{\perp}^2+m_D^2)t_M}{Qz_c}\Big]+\frac{1}{2}F\Big[\frac{(\Big[\vec{q}_{\perp}+\vec{k}_{\perp}\Big]^2-q_{\perp}^2)t_M}{Qz_c}\Big]
\Bigg\}\Bigg]\nn\\
&-&2g^2N_Ct_M\frac{\delta^{AB}}{2} \int \frac{d^{2}q_{\perp}}{(2\pi)^3}\frac{\delta^2(\vec{q}_{Tn}-\vec{q}_{\perp}-\vec{k}_{\perp})}{(\vec{q}_{\perp}+\vec{k}_{\perp})^2} \Bigg\{\left(\frac{\nu}{Qz_c}\right)^{\eta}\frac{1}{\eta}-F\Big[\frac{([\vec{q}_{\perp}+\vec{k}_{\perp}]^2+m_D^2)t_M}{Qz_c}\Big]
\Bigg\}
\eea
\normalsize
The piece proportional to $C_F$ gets canceled out by the vacuum jet function in the denominator so we ignore it here.
The first two lines give us the required piece for the BFKL evolution.  We will see that the last two lines will eventually cancel out with terms from $J_{n,V}$.


\item{Virtual Gluon diagrams $\mathcal{J}_{n,R}^V$}

We next turn to the virtual gluon radiative corrections.  We classify the diagrams in the same manner as the real gluon emissions.
All the diagrams that give a rapidity divergence are shown in Fig.\ref{VGW}.
   
\begin{figure}
\centering
\includegraphics[width=\linewidth]{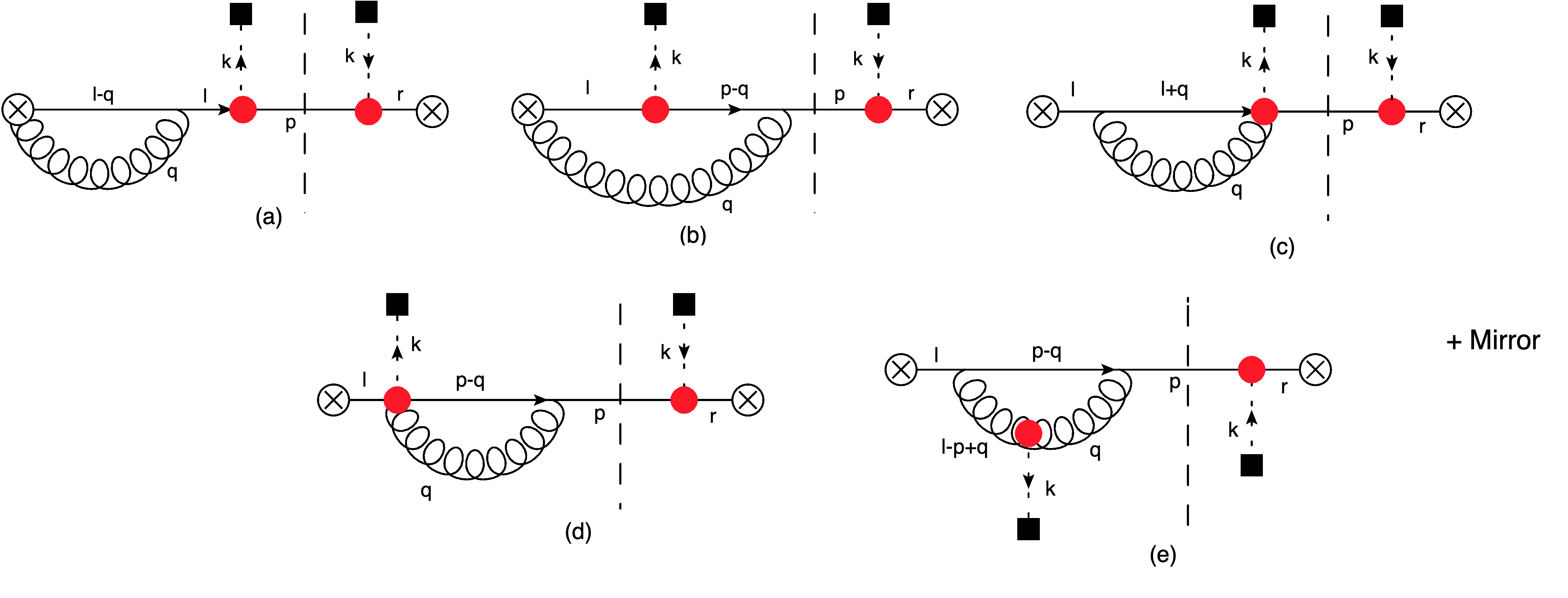}
\caption{Virtual Gluon emission diagrams with rapidity divergence }
\label{VGW}
\end{figure}      
  
Diagrams (a) and (b) along with their mirrors  add up to 
\bea
6(a)+6(b)&=&g^2t_M\delta^{AB}\int \frac{d^2q_{\perp}}{(2\pi)^{3}} \frac{\delta^2(\vec{q}_{Tn}-\vec{k}_{\perp}) }{q_{\perp}^2+m_D^2}\Bigg\{C_F\left(\frac{\nu}{Q}\right)^{\eta}\frac{1}{\eta}-\frac{N_C}{2}F\Big[\frac{(\vec{q}^2_{\perp}+m_D^2)t_M}{Q}\Big]\nn
\Bigg\}
\eea
Once again the piece proportional to $C_F$ will be divided out,  leaving behind only the $N_C$ term. 
 Next consider the contribution from Wilson line of the Glauber vertex from diagrams (c) and(d).   
Again we have two diagrams,  which add up to 
\bea
6(c)+6(d)&=&g^2t_M N_C\delta^{AB}\int \frac{d^2q_{\perp}}{(2\pi)^{3}} \frac{\delta^2(\vec{q}_{Tn}-\vec{k}_{\perp})}{q_{\perp}^2+m_D^2}\Bigg\{\left(\frac{\nu}{Q}\right)^{\eta}\frac{1}{\eta}-\frac{1}{2}F\Big[\frac{(\vec{q}^2_{\perp}+m_D^2)t_M}{Q}\Big]\Bigg\}\nn
\eea

Finally we have the gluon Glauber vertex contribution in diagram (c).
\bea
6(c)&=&-g^2t_M N_C\frac{\delta^{AB}}{2}\delta^2(\vec{q}_{Tn}-\vec{k}_{\perp}) \int \frac{d^2q_{\perp}}{(2\pi)^{3}} \frac{2\vec{q}_{\perp}\cdot (\vec{q}_{\perp}+\vec{k}_{\perp})}{(\vec{q}_{\perp}+\vec{k}_{\perp})^2+m_D^2}\frac{1}{\vec{q}_{\perp}^2+m_D^2}\nn\\
&&\Bigg\{\left(\frac{\nu}{Q}\right)^{\eta}\frac{1}{\eta}
-F\Big[\frac{([\vec{q}_{\perp}+\vec{k}_{\perp}]^2+m_D^2)t_M}{Q}\Big]\Bigg\}\nn
\eea

The result for all diagrams in Fig. \ref{VGW} now sum up to
\bea
 \mathcal{J}_{n,R}^V&=&g^2t_M N_C\frac{\delta^{AB}}{2}\int \frac{d^2q_{\perp}}{(2\pi)^{3}} \Bigg[\frac{k_{\perp}^2}{(\vec{q}_{\perp}+\vec{k}_{\perp})^2+m_D^2}\frac{\delta^2(\vec{q}_{Tn}-\vec{k}_{\perp}) }{\vec{q}_{\perp}^2+m_D^2}\Bigg\{\left(\frac{\nu}{Q}\right)^{\eta}\frac{1}{\eta}
-F\Big[\frac{([\vec{q}_{\perp}+\vec{k}_{\perp}]^2+m_D^2)t_M}{Q}\Big]\Bigg\}\nn\\
&+& \frac{\delta^2(\vec{q}_{Tn}-\vec{k}_{\perp})}{q_{\perp}^2+m_D^2}\Bigg\{F\Big[\frac{([\vec{q}_{\perp}+\vec{k}_{\perp}]^2+m_D^2)t_M}{Q}\Big]-F\Big[\frac{(\vec{q}^2_{\perp}+m_D^2)t_M}{Q}\Big]\Bigg\}\Bigg]
\eea  

\end{itemize}

 \subsection{$J_n^V$}	
 
 We can now look at the the part of the jet function which has Glauber vertex insertions on the same side of the cut.  This piece is defined as 
 \bea
  J_{n,V}^{AB}(e_n, \vec{b}, \vec{k}_{\perp}) = \frac{1}{k_{\perp}^2}\frac{\mathcal{J}_{n,V}^{AB}(e_n, \vec{b}, \vec{k}_{\perp})}{\mathcal{J}^{\perp}_n(e_n, \vec{b})}\nn
\eea
 where the vacuum jet function $\mathcal{J}^{\perp}_n(e_n, \vec{b})$ at one loop is given in Eq.\ref{eq:jetvac} while the numerator is defined in Eq.\ref{eq:jetv}.
 
\subsubsection{Tree level}

\begin{figure}
\centering
\includegraphics[width=0.5\linewidth]{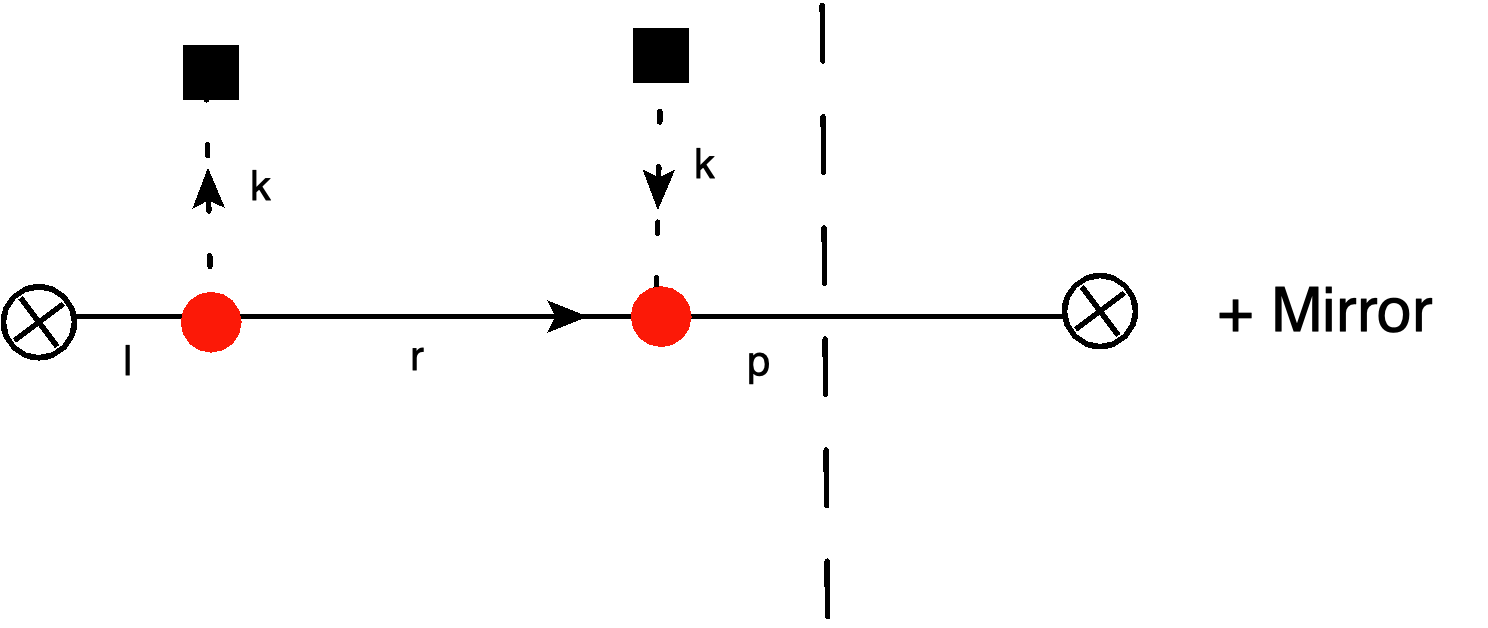}
\caption{Tree level diagram Glauber insertion on the same side of the cut.}
\label{JVT}
\end{figure}

The tree level diagram is shown in Fig. \ref{JVT}. This evaluates to 
\bea
\mathcal{J}_{n,V}^{AB(0)}&=&  \text{Tr}\Big[T^AT^B\Big]\int \frac{dl^+}{2\pi }\frac{ie^{-it_M/2l^+}}{l^+ +i\epsilon} \int \frac{dr^+}{2\pi}\frac{i}{r^+ - \frac{k_{\perp}^2}{Q}+i\epsilon}\text{Sinc}\Big[\frac{t_M}{2}l^+\Big]\delta^2(\vec{q}_{Tn})+c.c\nn
\eea

We can do the contour integral over $l^+$,  while doing the integral over $r^+$ directly which now gives us 
\bea
 \mathcal{J}_{n,V}^{AB(0)}&=& \frac{\delta^{AB}}{2}\delta^2(\vec{q}_{Tn})
\eea
\subsubsection{One loop}

Lets look at the real gluon emission first 

\begin{itemize}
\item{Real Gluon emission $\mathcal{J}_{n,V}^R$}\\
 First consider the diagrams which contain the Wilson line contribution from the hard vertex for a real gluon as shown in Fig.\ref{WJV}.  These, along with the mirror diagrams add up to 
 \begin{figure}
\centering
\includegraphics[width=\linewidth]{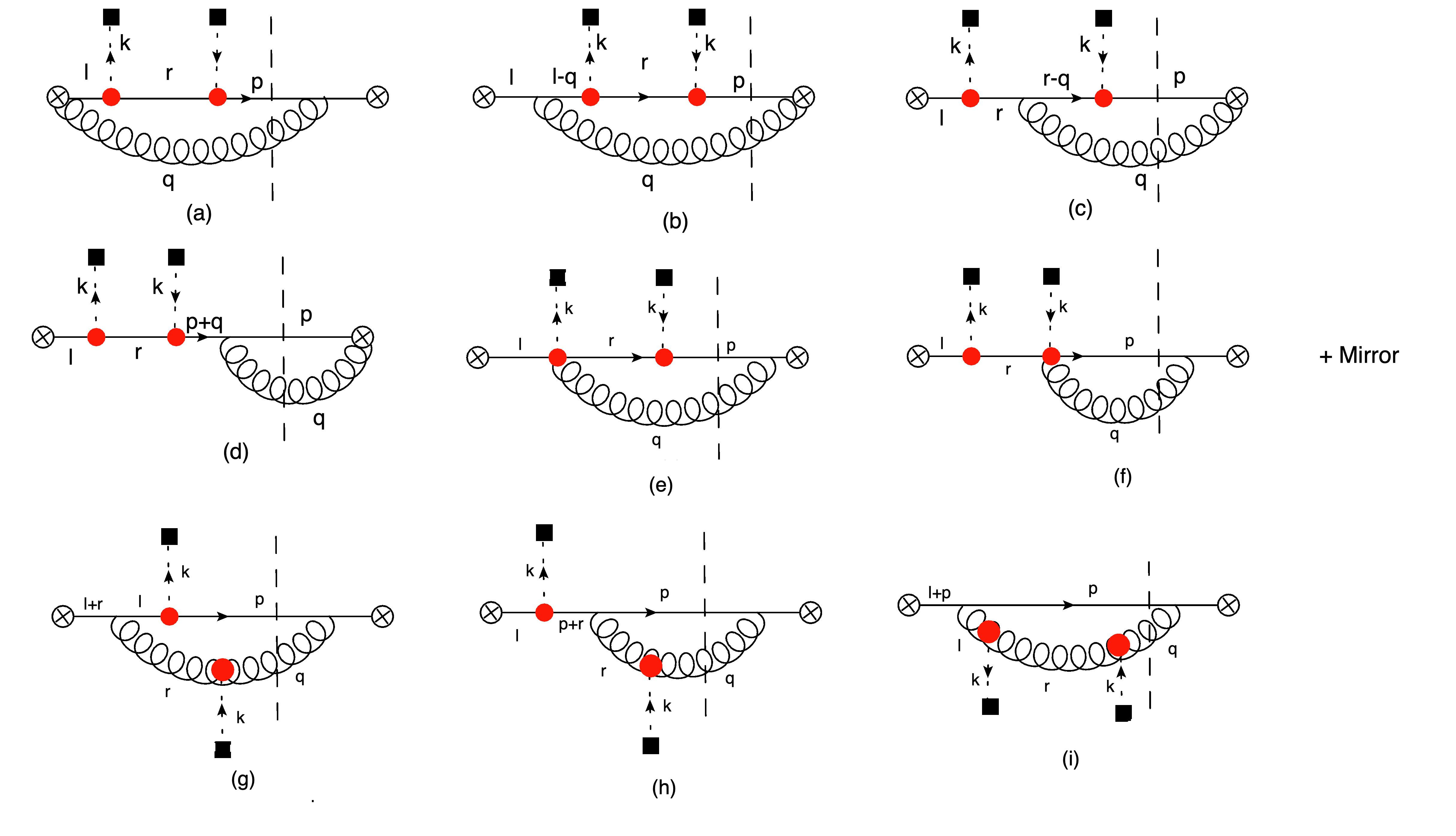}
\caption{Real Gluon emission for Glauber insertion on the same side of the cut.}
\label{WJV}
\end{figure}
\bea
 8(a)+8(b)+8(c)+8(d)&=&-g^2t_MC_F\frac{\delta^{AB}}{2}\int \frac{d^2q_{\perp}}{(2\pi)^{3}} \frac{\delta^2(\vec{q}_{Tn}-\vec{q}_{\perp}) }{q_{\perp}^2+m_D^2}\left(\frac{\nu}{Qz_c}\right)^{\eta}\frac{1}{\eta}
\eea
   This will again get divided out by the vacuum jet function. 
The next set of non-zero diagrams are ones with Wilson line real gluon from the Glauber vertex as shown in (e) and (f) which add up to 0.
  \bea 
8(e)+8(f) = 0 
  \eea
 
Next we consider the diagrams with one Gluon Glauber vertex and one quark Glauber vertex.  These are shown in diagrams (g) and (h).
\bea
8(g)+8(h) &=&  g^2 t_MN_C\frac{\delta^{AB}}{2}\int \frac{d^2q_{\perp}}{(2\pi)^3}\frac{2\vec{q}_{\perp} \cdot (\vec{q}_{\perp}+\vec{k}_{\perp})}{\left(q_{\perp}^2+m_D^2\right)\left(\vec{q}_{\perp}+\vec{k}_{\perp})^2+m_D^2\right)}\nn\\
&&\Bigg\{-F\Big[\frac{(\vec{q}_{\perp}+m_D^2)t_M}{Qz_c}\Big]+F\Big[\frac{[(\vec{q}_{\perp}+\vec{k}_{\perp})^2-q_{\perp}^2]t_M}{Qz_c}\Big]
\Bigg\}\delta^2(\vec{q}_{Tn}-\vec{q}_{\perp})\nn
   \eea

Finally we have the double gluon Glauber vertex diagram (i) which evalutes to 
 \bea
    8(i)&=&   -2g^2 t_M N_C \frac{\delta^{AB}}{2}\int \frac{d^2q_{\perp}}{(2\pi)^3}\frac{1}{q_{\perp}^2+m_D^2}\Bigg\{\left(\frac{\nu}{Qz_c}\right)^{\eta}\frac{1}{\eta}-F\Big[\frac{(\vec{q}_{\perp}^2+m_D^2)t_M}{Qz_c}\Big]\Bigg\}
   \eea

The net result for real gluon corrections here is simply 
\bea 
J_{n,V}^R&=& 8(g)+8(h)+8(i) 
\eea
We will see that the rapidity divergence will cancel out with part of $\mathcal{J}_{n,R}^R$,  leaving behind  a finite non-logarithmic contribution.

   
   \item{Virtual gluon emission $\mathcal{J}_{n,V}^V$}
   
We now consider the final set of diagrams that involve a virtual gluon contribution for Glauber insertions on the same side of the cut.  
First consider the diagrams with Wilson line from the hard vertex in Fig.\ref{WG2V}. These along with the complex conjugates add up to 
\bea
   9(a)+9(b)+9(c)+9(d)&=&2g^2t_M C_F\frac{\delta^{AB}}{2} \delta^2(\vec{q}_{Tn})\int \frac{d^2q_{\perp}}{(2\pi)^{3}} \frac{1}{q_{\perp}^2+m_D^2}\left(\frac{\nu}{Q}\right)^{\eta}\frac{1}{\eta}
 \eea
\begin{figure}
\centering
\includegraphics[width=\linewidth]{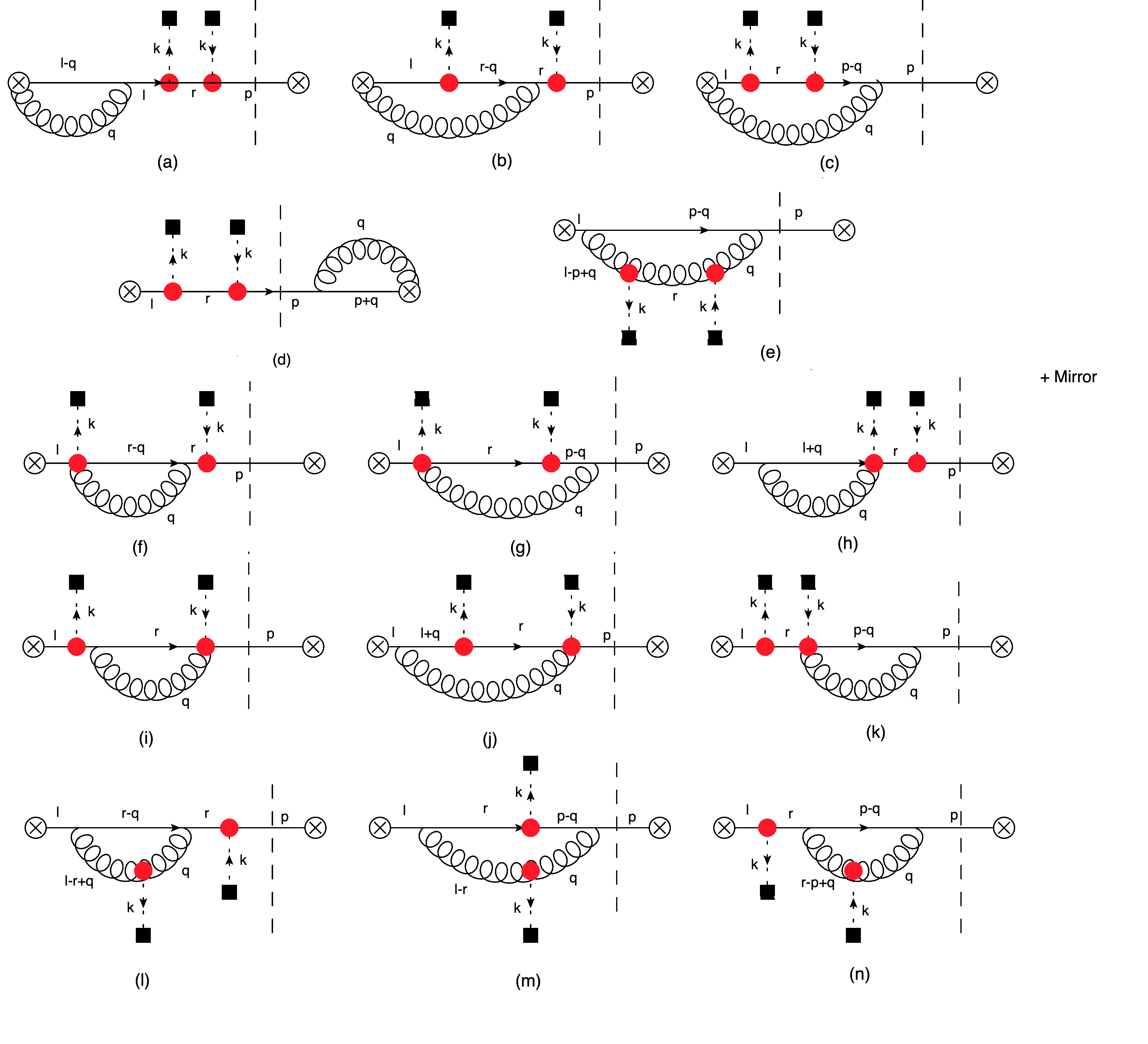}
\caption{Virtual gluon emission diagrams with rapidity divergence for Glauber insertion on the same side of the cut}
\label{WG2V}
\end{figure}         
Next we consider diagram(e) which evaluates to    
\bea
 9(e)&=&-2g^2t_MN_C\frac{\delta^{AB}}{2}\int \frac{d^2q_{\perp}}{(2\pi)^3}\frac{\delta^2(\vec{q}_{Tn})}{q_{\perp}^2+m_D^2}\Bigg\{\left(\frac{\nu}{Q}\right)^{\eta}\frac{1}{\eta}-F\Big[t_M\frac{q_{\perp}^2+m_D^2}{Q}\Big]\Bigg\}\nn
\eea      
   
Next we have the contribution from the wilson line of the glauber vertex. All of these pieces add up to 0.
\bea 
9(f)+9(g)+.. +9(k) = 0 
\eea

Next we consider the diagrams with one glauber gluon vertex. These add up to 
  \bea
 9(l)+19(m)+9(n)&=& g^2t_MN_C\frac{\delta^{AB}}{2}\int \frac{d^2q_{\perp}}{(2\pi)^3}\frac{2\vec{q}_{\perp}\cdot (\vec{q}_{\perp}+\vec{k}_{\perp})}{q_{\perp}^2+m_D^2}\frac{\delta^2(\vec{q}_{Tn})}{(\vec{q}_{\perp}+\vec{k}_{\perp})^2}\Bigg\{\left(\frac{\nu}{Q}\right)^{\eta}\frac{1}{\eta}\nn\\
 &-&F\Big[t_M\frac{(\vec{q}_{\perp}+\vec{k}_{\perp})^2}{Q}\Big]\Bigg\}
 \eea
  
We can combine these pieces, dividing out the vacuum jet function to write 
\bea
\mathcal{J}_{n,V}^{V}&=&  -g^2t_MN_C\frac{\delta^{AB}}{2}\int \frac{d^2q_{\perp}}{(2\pi)^3}\Bigg[\frac{k_{\perp}^2}{q_{\perp}^2+m_D^2}\frac{\delta^2(\vec{q}_{Tn})}{(\vec{q}_{\perp}+\vec{k}_{\perp})^2}\Bigg\{\left(\frac{\nu}{Q}\right)^{\eta}\frac{1}{\eta}-F\Big[t_M\frac{(\vec{q}_{\perp}+\vec{k}_{\perp})^2}{Q}\Big]\Bigg\}\nn\\
&+&\frac{\delta^2(\vec{q}_{Tn})}{q_{\perp}^2+m_D^2}\Bigg\{F\Big[t_M\frac{(\vec{q}_{\perp}+\vec{k}_{\perp})^2+m_D^2}{Q}\Big]-F\Big[t_M\frac{q_{\perp}^2+m_D^2}{Q}\Big]\Bigg\}\Bigg]
\eea

\end{itemize}
  
 \subsection{Full result and anomalous dimension }
In this section we add up all the pieces and present the full result for the jet function at one loop.  All the contributions in the previous sections are written in terms of a rapidity divergence $1/\eta$ that yields a logarithmic contribution of either $\ln \nu/Qz_c$ or $\ln \nu/Q$ along with a function F whose argument is the ratio of the medium lifetime $t_M$ and the formation time of the jet $t_F \sim Q/[q_{T}^2+m_D^2]$.  
The tree level jet function is 
\bea
 J^{AB(0)}_{\text{Med}}(\vec{k}_{\perp}, \vec{q}_{Tn})& = & \frac{t_M}{k_{\perp}^2}\frac{\delta^{AB}}{2}\Big\{\delta^2(\vec{q}_{Tn}-\vec{k}_{\perp})-\delta^2(\vec{q}_{Tn})\Big\}
\eea

We can write the full result in terms of a BFKL kernel and finite non-logarithmic corrections.
\bea
 J^{AB(1)}_{\text{Med}}&\equiv&  J^{AB(1)}_{\text{Med, BFKL}}+ J^{AB(1)}_{\text{Med,  fin}}
\eea
\small
\bea
J^{AB(1)}_{\text{Med, BFKL}}(\vec{k}_{\perp}, \vec{q}_{Tn})&=& - 2g^2N_C\frac{t_M}{k_{\perp}^2}\frac{\delta^{AB}}{2}\int \frac{d^{2}q_{\perp}}{(2\pi)^3}\frac{\delta^2(\vec{q}_{Tn}-\vec{q}_{\perp})}{(\vec{q}_{\perp}-\vec{k}_{\perp})^2+m_D^2}\frac{k_{\perp}^2}{\vec{q}_{\perp}^2} \Bigg\{\ln\frac{\nu}{Qz_c}-F\Big[\frac{(\vec{q}_{\perp}^2+m_D^2)t_M}{Qz_c}\Big]\Bigg\}\nn\\
&+ &2g^2N_C \frac{t_M}{k_{\perp}^2}\frac{\delta^{AB}}{2}\int \frac{d^{2}q_{\perp}}{(2\pi)^3}\frac{\delta^2(\vec{q}_{Tn})}{(\vec{q}_{\perp}-\vec{k}_{\perp})^2+m_D^2}\frac{k_{\perp}^2}{\vec{q}_{\perp}^2} \Bigg\{\ln\frac{\nu}{Q}-F\Big[\frac{(\vec{q}_{\perp}^2+m_D^2)t_M}{Q}\Big]\Bigg\}\nn\\
&+& g^2N_C\int \frac{d^{2}q_{\perp}}{(2\pi)^3}\frac{J^{AB(0)}_{\text{Med}}(\vec{k}_{\perp}, \vec{q}_{Tn})}{(\vec{q}_{\perp}-\vec{k}_{\perp})^2+m_D^2}\frac{k_{\perp}^2}{\vec{q}_{\perp}^2} \Bigg\{\ln\frac{\nu}{Q}-F\Big[\frac{(\vec{q}_{\perp}^2+m_D^2)t_M}{Q}\Big]\Bigg\}
\label{eq:JetBFKL}
\eea
\normalsize

and
\small
\bea
&&J^{AB(1)}_{\text{Med, fin}}(\vec{k}_{\perp}, \vec{q}_{Tn})=g^2N_C\frac{t_M}{k_{\perp}^2}\delta^{AB} \int \frac{d^{2}q_{\perp}}{(2\pi)^3}\frac{\delta^2(\vec{q}_{Tn}-\vec{q}_{\perp})}{(\vec{q}_{\perp}-\vec{k}_{\perp})^2+m_D^2}\frac{\vec{q}_{\perp}\cdot \vec{k}_{\perp}}{\vec{q}_{\perp}^2}\nn\\
&&\Bigg\{F\Big[t_M\frac{(\vec{q}_{\perp}-\vec{k}_{\perp})^2+m_D^2}{Qz_c}\Big]-F\Big[t_M\frac{q_{\perp}^2+m_D^2}{Qz_c}\Big]\Bigg\}\nn\\
&-&g^2\frac{N_C}{2}\frac{t_M}{k_{\perp}^2}\delta^{AB} \int \frac{d^{2}q_{\perp}}{(2\pi)^3}\Bigg\{F\Big[t_M\frac{(\vec{q}_{\perp}-\vec{k}_{\perp})^2+m_D^2}{Q}\Big]-F\Big[t_M\frac{q_{\perp}^2+m_D^2}{Q}\Big]\Bigg\}\frac{\delta^2(\vec{q}_{Tn}-\vec{k}_{\perp})+\delta^2(\vec{q}_{Tn})}{(\vec{q}_{\perp}-\vec{k}_{\perp})^2+m_D^2}
\eea
\normalsize
Given the definition of the function F in Eq.\ref{eq:F}., we can confirm that $J^{AB(1)}_{\text{Med, fin}}(\vec{k}_{\perp}, \vec{q}_{Tn})$ does not have any UV or rapidity divergences, nor does it contain large logarithmic corrections and therefore it only gives an O($\alpha_s$) correction. 

We can then write the rapidity renormalization group equation for the jet function as 
\bea
\nu \frac{d}{d \nu} J^{AB}_{\text{Med}} = -\frac{\alpha_s N_C}{\pi^2}\int \frac{d^2q_{\perp}}{(\vec{q}_{\perp}-\vec{k}_{\perp})^2}\Bigg\{J_{\text{Med}}^{AB}(\vec{q}_{\perp},\vec{q}_{Tn})-\frac{1}{2}\frac{J_{\text{Med}}^{(AB)}(\vec{k}_{\perp},\vec{q}_{Tn})k_{\perp}^2}{q_{\perp}^2}\Bigg\}
\eea
which is just the BFKL equation and is identical to the case of the long lived medium. This is not surprising since the medium structure function remains unchanged and we therefore expect no change in the renormalization group equation for the jet function..  We therefore see that the finite lifetime of the medium does not affect the anomalous dimension.  However,  as we will see below it does affect the natural rapidity scale since the function F in the Eq. \ref{eq:JetBFKL} can give a large contribution and modify the argument of the rapidity logarithm.

 We can consider three distinct hierarchies of time scales and simplify the result in each case 

\begin{enumerate}
\item{ $t_M \gg t_F$} \\
This would bring us back to the case of a long lived medium.  In this limit $F(x) \rightarrow \delta(x) \rightarrow 0$ . In this case,  jet function obeys a BFKL equation with a natural rapidity scale $\nu \sim Q \sim Qz_c$.  This was already considered in \cite{Vaidya:2021vxu} and we refer the reader to that paper for more details. 

\item{$t_M \sim t_F$}\\
In this case the function F yields a finite non logarithmic correction.  The jet function continues to have a BFKL evolution with a scale which is now $\nu \sim Qz_c \sim Q \sim t_Mq_{T}^2 $.

\item{$t_M \ll t_F$} 
This is the case of a short-lived medium for which $F(x) \rightarrow \ln x$, which gives a logarithmic enhancement.  While the anomalous dimension continues to be a BFKL kernel,  the natural rapidity scale for the jet function now changes to $\nu \sim q_T^2t_M$  from $\nu \sim Q \sim Qz_c$.  This intuitively makes sense as the logarithm resummed by solving the BFKL equation now changes from $\ln Q/k_{\perp}$ to  a smaller value $\ln q_T^2t_M/k_{\perp}$ since there is not enough time for the energetic partons of the jet to go on-shell before the medium disappears. 

\end{enumerate}
   
 \section{ RG evolution and cross section}
\label{sec:Resum} 
From the previous section, it is clear that the medium induced jet function still obeys the BFKL equation albeit with a modified scale when the medium lifetime is much shorter than the formation time of the jet. Therefore the resummation follows the same steps  as \cite{Vaidya:2021vxu} and is amenable to a numerical solution.  The only difference is that BFKL running will resum a $\ln t_M q_T^2/k_{\perp}$. Therefore, we will not discuss the details of resummation here, leaving the numerical result for a full phenomenological prediction in a future work.  
We can now look at our density matrix element to quadratic order in the Glauber expansion 
\bea
\Sigma(t,\vec{q}_T) = \Sigma^{(0)}(\vec{q}_T)+ \Sigma^{(2)}(\vec{q}_T)
\label{Sig}
\eea
where we have 
\bea
\Sigma^{(2)}(\vec{q}_T) = t_M|C_{G}|^2(Q)  \Sigma^{(0)}(Q,z_{c},\vec{q}_T)\otimes_{q_T}\int \frac{d^2k_{\perp}}{(2\pi)^2}S_G^{AB}(k)\mathcal{J}_{n,\text{Med}}^{AB}(Q, z_{c},\vec{q}_T,\vec{k}_{\perp})
\eea
$C_{G} = 8\pi \alpha_s(Q)$ is the Hard function.
We can use the RGE solutions described in the previous section to evolve medium induced soft and jet functions in rapidity and virtuality so that 
\bea
 \Sigma^{(2)}(\vec{q}_T) = t_M|C_{qq}|^2(Q)  \Sigma^{(0)}(Q,z_{c},\vec{q}_T)\otimes_{q_T}K_{\text{Med}}(\vec{q}_T, t)
\eea
with 
\bea
\label{KMed}
K_{\text{Med}}(p_{\perp}) =(N_c^2-1)\int \frac{d^2k_{\perp}}{(2\pi)^2} S_G^{\text{resum}}(k_{\perp})J^{\text{resum}}(Q, z_{c}, \vec{p}_{\perp}, k_{\perp}) 
\eea
where the resummed jet and soft functions are obtained by solving the BFKL equation and the running of the strong coupling as outlined in \cite{Vaidya:2021vxu}.
The density matrix element can now be related to the cross section writing 
\bea
\frac{d\sigma(t_M)}{d^2\vec{q}_T} = \mathcal{N}\frac{\Sigma(t_M)}{V}
\eea
where $\mathcal{N}$ is a simple kinematical factor while V is the 4d volume. 
Eq. \ref{Sig}  describes the evolution of the density matrix over a time scale t.  We can write the evolution equation in a suggestive form by going to impact parameter space( $\vec{b}$) which is conjugate to $\vec{q}_T$,
\bea
\Sigma(t_M,\vec{b}) = \Sigma^{(0)}(\vec{b})\Big[1+ \frac{t_M}{\lambda_{\text{mfp}}(\vec{b})}\Big]
\label{Evomfp}
\eea
where 
\bea
\lambda^{-1}_{\text{mfp}}(b) = |C_{qq}|^2K_{\text{Med}}(\vec{b})
\label{mfp}
\eea
can be thought of as  the inverse mean free path of the jet in the medium. If the medium is very dilute then the mean free path(mfp) can be large compared to $t_M$ in which case Eq.\ref{Evomfp} is a good enough approximation. 

The second case is when the medium is dense enough so that the m.f.p. is comparable to or smaller than the medium size then it becomes necessary to resum higher powers of $t/\lambda_{\text{mfp}}$. These higher order corrections correspond to multiple interactions of the jet with the medium. We are working in a hierarchy when $ 1/T \ll \lambda_{\text{mfp}} \leq t_M \leq t_F$.  To proceed, we need to consider higher order Glauber insertions and derive a factorization formula at each order in the Glauber expansion.  Since all time scales are still much greater than the typical formation time for the soft radiation (1/T), we expect that at 2n order in the Glauber exchange, the cross section will factorize into n copies of the Medium structure function. This will not be true for the medium induced jet function since quantum interference between successive interactions of the the jet with the medium(LPM effect) now becomes important. However, we still expect that by Renormalization Group consistency,  the medium jet function will  have a rapidity anomalous dimension n$\times$ BFKL.  We will consider this interesting scenario in an upcoming paper \cite{varun}.
  
 \section{Conclusion} 
\label{sec:Conclusion}  
The purpose of this paper was to investigate the possibility that a factorization formula showing a manifest separation of UV and IR scales can be derived for a generic jet substructure observable even in the case of a short lived medium created during heavy ion collisions. I have shown that such a factorization formula can be derived in terms of a medium structure function or "PDF of the medium" that encodes the observable independent soft physics of the QGP and the medium induced jet function under certain assumptions 
\begin{enumerate}
\item{} The lifetime of the medium $t_M$ is much longer than $t_c \sim$ 1/T, i.e the relaxation time in the QGP bath while being of the order of or smaller than the formation time of the jet $t_F\sim Q/q_T^2$. This is a phenomenologically reasonable assumption since $t_M$ is usually an order of magnitude larger than $t_c$.  This implies that the soft physics at the scale $T$  is not sensitive to $t_M$ and enjoys the same dynamics as in the long lived medium, thus establishing its universal nature. 
\item{} The QGP medium created is homogeneous in space and over its lifetime. While this is a idealized scenario, it is not a necessary  condition for factorization. Since a single interaction of the jet with the medium is virtually instantaneous in time and  probes it over length scales inverse of the transverse momentum exchanged $\sim 1/k_{\perp}$, the factorization is robust against inhomogeneties in the medium over length scales $\gg 1/k_{\perp} \sim 1/T$, i.e. as long as the jet sees a uniform medium over the length-time scales of a single Glauber exchange. The factorization formula incorporating such inhomogeneity will be  considered in an upcoming paper \cite{varun}. 
\item{} The medium is dilute, i.e the mean free path of the jet in the medium($\lambda_{\text{mfp}}$) is much smaller than the medium lifetime $t_M$. Again this is not necessary for deriving factorization but it simply allows us to stop at the level of a single jet-medium interaction. To extend the current formalism for the case of the dense medium then requires us to systematically incorporate higher order Glauber exchanges, write a factorization formula at each order and resum the  dominant corrections. This will also include the LPM effect which encodes quantum interference between successive interactions with the medium and will be the subject of an upcoming paper \cite{varun}. 
\end{enumerate}
  
The medium structure function is identical to that for a long lived medium considered in \cite{Vaidya:2021vxu}. Therefore, it continues to obey the BFKL evolution equation in rapidity and the QCD running in virtuality with a natural scale $k_{\perp}$.  The medium jet function is modified, in particular the "+" component of momentum, which is the small light cone momentum component of the jet, is not conserved in jet-medium interaction due to the short medium lifetime. This also means that the energetic partons of the jet created during the hard interaction $do$ $not$ go on-shell before interacting with the medium. Therefore, quantum interference between the hard vertex that creates the jet and subsequent medium interactions is now important. The consequence of these modifications is to induce additional finite corrections at one loop that modifies the natural rapidity scale from the hard scale Q to a smaller value $q_T^2t_M$ while still obeying the BFKL equation. 
\acknowledgments
I thank Yuri Kovchegov for pointing out the phenomenological relevance of a short lived medium in heavy ion collisions. This work is supported by the Office of Nuclear Physics of the U.S. Department of Energy under Contract DE-SC0011090 and Department of Physics, Massachusetts Institute of Technology.

\appendix
\section{One loop medium jet function}
\label{app:OneLoop}
In this section I provide the mathematical formulae for computing the one loop result for the medium jet function.  I will provide the integrands for all the diagrams for computing $\mathcal{J}_{n,R}$ while evaluating a few in detail. 

\subsection{Real Gluon emission for $\mathcal{J}_{n,R}$}
We start with the diagram shown in Fig.\ref{JR}(a). We will evaluate this diagram in detail.  We will only keep the dominant logarithmic contributions while ignoring finite O($\alpha_s$) pieces. The final results are already given in the main body of the paper. At the Glauber vertex, we can implement momentum conservation along the $\perp$ and $-$ directions but not the $+$.  
Since these diagram have rapidity divergences, we introduce an $\eta$ regulator along with a renormalization scale $\nu$\cite{Chiu:2012ir} along with usual dimensional regularization. This is because dim. reg. is boost invariant and therefore cannot regulate the divergence from the separation of soft physics from the collinear since both modes have the same virtuality.
\small
\bea
2(a)&=& 4g^2\text{Tr}[T^CT^AT^BT^C]t_M \nu^{\eta}\int d^4p p^-\delta^+(p^2)\int \frac{d^dq}{(2\pi)^{d-1}}\frac{|q^-|^{-\eta}}{q^-}\delta^+(q^2-m_D^2)\nn\\
&&\int \frac{dl^+}{2\pi}\frac{i}{\Big[l^+ -\frac{(\vec{p}_{\perp}+\vec{k}_{\perp})^2}{p^-}+i\epsilon\Big]} \int \frac{dr^+}{2\pi}\frac{(-i)}{\Big[r^+- \frac{(\vec{p}_{\perp}+\vec{k}_{\perp})^2}{p^-}-i\epsilon\Big]} \frac{e^{-i\frac{t_M}{2}(l^+-r^+)}}{\Big[r^++q^+ - \frac{\left(\vec{p}_{\perp}+ \vec{q}_{\perp}+\vec{k}_{\perp}\right)^2}{p^-}-i\epsilon \Big]}\nn\\
&&\text{Sinc} \Big[(l^+-r^+)\frac{t_M}{2}\Big]\delta^2(\vec{p}_{\perp}+\vec{k}_{\perp}+\vec{q}_{\perp})\delta(Q-p^--q^-)\Theta(Qz_c-q^-)\delta^2(\vec{q}_{Tn}-\vec{q}_{\perp}-\vec{k}_{\perp})\nn\\
&=&  g^2\frac{C_F}{2}\delta^{AB}t_M \nu^{\eta}\int \frac{d^{2-2\epsilon}q_{\perp}\int_0^{Qz_c}dq^-}{(2\pi)^{d-1}}\frac{|q^-|^{-\eta}}{(q^-)^2}\delta^2(\vec{q}_{Tn}-\vec{q}_{\perp}-\vec{k}_{\perp})\nn\\
&&\int \frac{dl^+}{2\pi}\frac{i}{\Big[l^+ -\frac{q_{\perp}^2}{Q-q^-}+i\epsilon\Big]} \int \frac{dr^+}{2\pi}\frac{(-i)}{\Big[r^+- \frac{q_{\perp}^2}{Q-q^-}-i\epsilon\Big]} \frac{e^{-i\frac{t_M}{2}(l^+-r^+)}}{\Big[r^++\frac{q_{\perp}^2}{q^-}-i\epsilon \Big]}\text{Sinc} \Big[(l^+-r^+)\frac{t_M}{2}\Big]\nn
\eea
\normalsize
We can now do the contour integral over $l^+$ closing the contour in the lower half complex plane,
\small
\bea
2(a)&=&  g^2\frac{C_F}{2}\delta^{AB}t_M \nu^{\eta}\int \frac{d^{2-2\epsilon}q_{\perp}\int_0^{Qz_c}dq^-}{(2\pi)^{d-1}}\frac{|q^-|^{-\eta}}{(q^-)^2}\int \frac{dr^+}{2\pi}\frac{1}{\Big[r^+- \frac{q_{\perp}^2}{Q-q^-}-i\epsilon\Big]} \frac{e^{-i\frac{t_M}{2}(\frac{q_{\perp}^2}{Q-q^-}-r^+)}}{\Big[r^++\frac{q_{\perp}^2+m_D^2}{q^-}-i\epsilon \Big]}\nn\\
 &&\text{Sinc} \Big[(\frac{q_{\perp}^2}{Q-q^-}-r^+)\frac{t_M}{2}\Big]\delta^2(\vec{q}_{Tn}-\vec{q}_{\perp}-\vec{k}_{\perp})
\eea
\normalsize
Now we can do the contour integral over $r^+$ closing the contour in the upper half plane.  We have two residues 
\begin{enumerate}
\item{}
\bea
2(a)_1&=&  g^2\frac{C_F}{2}\delta^{AB}t_M \nu^{\eta}\int \frac{d^{2-2\epsilon}q_{\perp}dq^-}{(2\pi)^{d-1}}\frac{|q^-|^{-\eta}}{q^-}\frac{Q-q^-}{Qq_{\perp}^2+m_D^2(Q-q^-)}\delta^2(\vec{q}_{Tn}-\vec{q}_{\perp}-\vec{k}_{\perp})\nn 
\eea
This is identical to the contribution that was already captured in the case of a long lived medium and has a rapidity divergence. This now leads to the rapidity logarithm (taking the limit $q^- \rightarrow 0$) and ignoring any finite non-logarithmic contributions
\small
\bea
 2(a)_1&=&  -g^2\frac{C_F}{2}\delta^{AB}t_M \left(\frac{\nu}{Qz_c}\right)^{\eta}\frac{1}{\eta}\int \frac{d^{2-2\epsilon}q_{\perp}}{(2\pi)^{d-1}}\frac{1}{q_{\perp}^2+m_D^2}\delta^2(\vec{q}_{Tn}-\vec{q}_{\perp}-\vec{k}_{\perp})\nn
\eea
\normalsize
\item{}
The second pole in $r^+$ gives a residue which is a new contribution,
\small
\bea
2(a)_2&=&  g^2\frac{C_F}{2}\delta^{AB}t_M \nu^{\eta}\int \frac{d^{2-2\epsilon}q_{\perp}\int_0^{Qz_c}dq^-}{(2\pi)^{d-1}}\frac{|q^-|^{-\eta}}{(q^-)^2}\frac{e^{-i\frac{t_M}{2}(\frac{q_{\perp}^2}{Q-q^-}+\frac{q_{\perp}^2+m_D^2}{q^-})}}{\Big[-\frac{q_{\perp}^2+m_D^2}{q^-}- \frac{q_{\perp}^2}{Q-q^-}-i\epsilon\Big]}\nn\\
&&\text{Sinc} \Big[(\frac{q_{\perp}^2}{Q-q^-}+\frac{q_{\perp}^2+m_D^2}{q^-})\frac{t_M}{2}\Big]\delta^2(\vec{q}_{Tn}-\vec{q}_{\perp}-\vec{k}_{\perp})
\eea
\normalsize
To get a real answer, we also need to add in the complex conjugate from the mirror diagram which will double the result from the first pole and for the second pole we have  
\small
\bea
 2(a)_2&=&  g^2\frac{C_F}{2}\delta^{AB}t_M \nu^{\eta}\int \frac{d^{2-2\epsilon}q_{\perp}\int_0^Qdq^-}{(2\pi)^{d-1}}\frac{|q^-|^{-\eta}}{(q^-)^2}\frac{2\text{Cos} \Big[\frac{t_M}{2}(\frac{q_{\perp}^2}{Q-q^-}+\frac{q_{\perp}^2+m_D^2}{q^-})\Big]}{\Big[-\frac{q_{\perp}^2+m_D^2}{q^-}- \frac{q_{\perp}^2}{Q-q^-}-i\epsilon\Big]}\nn\\
 &&\text{Sinc} \Big[(\frac{q_{\perp}^2}{Q-q^-}+\frac{q_{\perp}^2+m_D^2}{q^-})\frac{t_M}{2}\Big]\delta^2(\vec{q}_{Tn}-\vec{q}_{\perp}-\vec{k}_{\perp})
\eea
\normalsize
Since we are interested in the divergence that happens as $q^- \rightarrow 0$, we will take this limit which is same as expanding the integrand around $q^-=0$ thus dropping any non-logarithmic contributions. This simplifies to 
\small
\bea
2(a)_2&=& - 2g^2\frac{C_F}{2}\delta^{AB}\int \frac{d^{2-2\epsilon}q_{\perp}\int_0^{Qz_c}dq^-}{(2\pi)^{d-1}}\frac{\text{Sin} \Big[t_M(\frac{q_{\perp}^2+m_D^2}{q^-})\Big]}{\Big[q_{\perp}^2+m_D^2\Big]^2}\delta^2(\vec{q}_{Tn}-\vec{q}_{\perp}-\vec{k}_{\perp})\nn
\eea
\normalsize
which evaluates to 
\small
\bea
2(a)_2&=&  - 2g^2\frac{C_F}{2}\delta^{AB}\int \frac{d^{2-2\epsilon}q_{\perp}}{(2\pi)^{d-1}\Big[q_{\perp}^2+m_D^2\Big]^2}\Bigg\{-(q_{\perp}^2+m_D^2)t_M \text{CosInt}\Big[\frac{(q_{\perp}^2+m_D^2)t_M}{Qz_c}\Big]\nn\\
 &+&Qz_c\text{Sin}\Big[\frac{(q_{\perp}^2+m_D^2)t_M}{Qz_c}\Big]\Bigg\}\delta^2(\vec{q}_{Tn}-\vec{q}_{\perp}-\vec{k}_{\perp})
\eea
\normalsize
written in terms of the Cosine integral 
\bea
\text{ CosInt}(x) = \int_x^{\infty} \frac{ dz \text{Cos} z}{z}
\eea
 
\end{enumerate}
Adding up the contributions from the two poles(and including the mirror diagrams)  we have 
\small
\bea
2(a)&=&   -g^2C_F\delta^{AB}t_M \int \frac{d^{2}q_{\perp}}{(2\pi)^{3}}\frac{\delta^2(\vec{q}_{Tn}-\vec{q}_{\perp}-\vec{k}_{\perp})}{q_{\perp}^2+m_D^2}\Bigg\{\left(\frac{\nu}{Qz_c}\right)^{\eta}\frac{1}{\eta}-\text{CosInt}\Big[\frac{(q_{\perp}^2+m_D^2)t_M}{Qz_c}\Big]\nn\\
&+&\text{Sinc}\Big[\frac{(q_{\perp}^2+m_D^2)t_M}{Qz_c}\Big]\Bigg\}\nn
\eea
\normalsize
We see  that in the case $t_M  \sim Qz_c/q_{\perp}^2$, i.e. the size of the medium is comparable to the formation time of the jet,  we get an O($\alpha_s$) finite contribution.  On the other hand, in the case where $t_F^c \gg  t_M$ we can expand the result in the limit $\frac{(q_{\perp}^2+m_D^2)t_M}{Qz_c} \ll 1$, which leads to 
\small
\bea
2(a)&=&  -2g^2\frac{C_F}{2}\delta^{AB}t_M\int \frac{d^{2}q_{\perp}\delta^2(\vec{q}_{Tn}-\vec{q}_{\perp}-\vec{k}_{\perp})}{(2\pi)^{3}\Big[q_{\perp}^2+m_D^2\Big]}\Bigg\{\frac{1}{\eta}+\ln\frac{\nu}{t_M(q_{\perp}^2+m_D^2)}+1\Bigg\}
\eea
\normalsize
So the net effect on this diagram has been to change the natural scale for rapidity from $\nu \sim Qz_c $ to $\nu \sim t_M(q_{\perp}^2+m_D^2)$.

However,  as we shall see after adding up all the diagrams,  the corrections proportional to $C_F$ which correspond to a vacuum evolution will in fact revert back to a natural rapidity scale of $\nu \sim Qz_c$.  Only the medium induced corrections proportional to $N_C$  will see this modification of the rapidity scale.
This is physically intuitive since we should not expect the vacuum shower to be sensitive to the lifetime of the medium.

The other diagram with quark Glauber vertices and Wilson line contribution from the hard vertex shown in Fig.\ref{JR}(b). This represents the interference between the hard vertex and the Glauber vertex.
\small
\bea
2(b)&=&4t_M g^2\text{Tr}[T^CT^AT^CT^B]\nu^{\eta}\int d^4p \delta^+(p^2)p^-\int \frac{d^4 q|q^-|^{-\eta}}{(2\pi)^3q ^-}\delta^+(q^2-m_D^2)\nn\\
&&\int \frac{dl^+}{2\pi}\frac{i e^{-i\frac{t_M}{2}(l^+-p^+)}}{l^+-\frac{(\vec{p}_{\perp}+\vec{k}_{\perp})^2}{p^-}+i\epsilon}\int \frac{dr^+}{2\pi}\frac{(-i) e^{-it_M/2(p^++q^+-r^+)}}{r^+-i\epsilon}\frac{\delta^2(\vec{p}_{\perp}+\vec{k}_{\perp}+\vec{q}_{\perp})\delta(Q-p^--q^-)}{(p+q)^2-i\epsilon}\nn\\
&&\text{Sinc}\Big[\frac{t_M}{2}(l^++q^+-r^+)\Big]\Theta(Qz_c-q^-)\delta^2(\vec{q}_{Tn}-\vec{q}_{\perp}-\vec{k}_{\perp})
\eea
\normalsize

We can  add the mirror diagram and expand around $q^- =0$, which leads to 
\small
\bea
  2(b)&=&t_M g^2\delta^{AB}\Big[\frac{N_c}{2}-C_F\Big]\int \frac{d^{2}q_{\perp}}{(2\pi)^{3}}\frac{\delta^2(\vec{q}_{Tn}-\vec{q}_{\perp}-\vec{k}_{\perp})}{q_{\perp}^2+m_D^2}\Bigg\{\text{CosInt}\Big[\frac{(q_{\perp}^2+m_D^2)t_M}{Qz_c}\Big]\nn\\
  &-&\text{Sinc}\Big[\frac{(q_{\perp}^2+m_D^2)t_M}{Qz_c}\Big]\Bigg\}
\eea
\normalsize
Adding in 2(a), we see that for the $C_F$ contribution,  the natural rapidity scale comes back to $\nu \sim Qz_c$. 

We now give the integrands for the rest of the real gluon emission diagrams. For Fig.\ref{djgw},
\small
\bea
4(a)&=& 4g^2t_M\nu^{\eta}\frac{N_c}{2}\frac{\delta^{AB}}{2}\int d^4p p^-\delta(p^2)\int \frac{d^4q|q^-|^{-\eta}}{(2\pi)^3q^-}\frac{\delta(q^2-m_D^2)(p^-+q^-)}{(p+q)^2-i\epsilon}\delta^2(\vec{p}_{\perp}+\vec{q}_{\perp}+\vec{k}_{\perp})\nn\\
&& \int \frac{dl^+}{2\pi}\frac{i e^{-i\frac{t_M}{2}(l^+-p^+-q^+)}}{l^+ -\frac{(\vec{p}_{\perp}+\vec{k}_{\perp}+\vec{q}_{\perp})^2}{p^-+q^-}+i\epsilon} \int \frac{dr^+}{2\pi}\frac{(-i)e^{-i\frac{t_M}{2}(p^++q^+-r^+)}}{r^+-\frac{(\vec{p}_{\perp}+\vec{k}_{\perp}+\vec{q}_{\perp})^2}{p^-+q^-}-i\epsilon}\text{Sinc}\Big[\frac{t_M}{2}(l^+-r^+)\Big]\nn\\
&&\delta^2(\vec{q}_{Tn}-\vec{q}_{\perp}-\vec{k}_{\perp})\delta(Q-p^--q^-)\Theta(Qz_c-q^-)\nn
\eea 
\normalsize

\bea
4(b)&=& -4g^2t_M\frac{N_c}{2}\frac{\delta^{AB}}{2}\int d^4p p^-\delta(p^2)\int \frac{d^4q}{(2\pi)^3q^-}\delta(q^2-m_D^2)(p^-+q-)\delta(Q-p^--q^-)\nn\\
&& \int \frac{dl^+}{2\pi}\frac{i e^{-i\frac{t_M}{2}(l^+-p^+-q^+)}}{l^+ -\frac{(\vec{p}_{\perp}+\vec{k}_{\perp}+\vec{q}_{\perp})^2}{p^-+q^-}+i\epsilon} \int \frac{dr^+}{2\pi} \frac{(-i)e^{-i\frac{t_M}{2}(p^+-r^+)}}{r^+-\frac{(\vec{p}_{\perp}+\vec{k}_{\perp})^2}{p^-}-i\epsilon}\frac{1}{\left(q^++r^+-\frac{(\vec{p}_{\perp}+\vec{k}_{\perp}+\vec{q}_{\perp})^2}{p^-+q^-}-i\epsilon\right)}\nn\\
&&\text{Sinc}\Big[\frac{t_M}{2}(l^+-r^+-q^+)\Big]\delta^2(\vec{q}_{Tn}-\vec{q}_{\perp}-\vec{k}_{\perp})\delta(Q-p^--q^-)\Theta(Qz_c-q^-)\nn
\eea
Fig. \ref{djgl} diagrams:
\small
\bea
5(a)&=& 4g^2\frac{N_C}{2}t_M\frac{\delta^{AB}}{2}\int d^4p \delta^+(p^2) \int \frac{d^4q}{(2\pi)^3} \delta^+(q^2-m_D^2)\delta^2(\vec{q}_{\perp}+\vec{p}_{\perp}+\vec{k}_{\perp})\delta(Q-p^--q^-)\nn\\
&&\int \frac{dl^+ i}{2\pi(l^2-m_D^2+i\epsilon)}\frac{ie^{-i\frac{t_M}{2}(l^+-q^+)}}{p^++l^+ +i\epsilon}\int \frac{dr^+}{2\pi}\frac{(-i) e^{i\frac{t_M}{2}(r^+-p^+-q^+)}}{r^+-i\epsilon}\text{Sinc}\Big[\frac{t_M}{2}(l^+-r^++p^+)\Big]\nn\\
&&\frac{Q}{(p+q)^2-i\epsilon}\frac{2\vec{q}_{\perp}\cdot (\vec{q}_{\perp}+\vec{k}_{\perp})}{q^-}\delta^2(\vec{q}_{Tn}-\vec{q}_{\perp}-\vec{k}_{\perp})\Theta(Qz_c-q^-)\nn
\eea
\normalsize

\small
\bea
5(b)&=&  -4g^2N_Ct_M\frac{\delta^{AB}}{2}\int d^4p \delta^+(p^2) \int d^4q \delta^+(q^2-m_D^2)\delta^2(\vec{q}_{\perp}+\vec{p}_{\perp}+\vec{k}_{\perp})\delta(Q-p^--q^-)\nn\\
&&\int \frac{dl^+}{(l^2-m_D^2+i\epsilon)}\frac{e^{-i\frac{t_M}{2}(l^+-q^+)}}{p^++l^+ +i\epsilon}\int \frac{dr^+ e^{-i\frac{t_M}{2}(p^+-r^+)}}{\left(r^+-\frac{(\vec{p}_{\perp}+\vec{k}_{\perp})^2}{p^-}-i\epsilon\right)\left(r^++q^+-i\epsilon\right)}\nn\\
&&\text{Sinc}\Big[\frac{t_M}{2}(l^+-r^++p^+-q^+)\Big]\Bigg[\frac{2\vec{q}_{\perp}\cdot (\vec{q}_{\perp}+\vec{k}_{\perp})}{q^-}\Bigg]\delta^2(\vec{q}_{Tn}-\vec{q}_{\perp}-\vec{k}_{\perp})\Theta(Qz_c-q^-)\nn
\eea

\bea
 5(c)&=&  2g^2N_Ct_M\frac{\delta^{AB}}{2}\int d^4p \delta^+(p^2) \int \frac{d^4q}{(2\pi)^3} \delta^+(q^2-m_D^2)\delta^2(\vec{q}_{\perp}+\vec{p}_{\perp}+\vec{k}_{\perp})\delta(Q-p^--q^-)\nn\\
&&\int \frac{dl^+}{(l^2-m_D^2+i\epsilon)}\frac{e^{-i\frac{t_M}{2}(l^+-q^+)}}{p^++l^+ +i\epsilon}\int \frac{dr^+ e^{-i\frac{t_M}{2}(q^+-r^+)}}{\left(r^2-m_D^2-i\epsilon\right)\left(r^++p^+-i\epsilon\right)}\text{Sinc}\Big[\frac{t_M}{2}(l^+-r^+)\Big]\nn\\
&&(\vec{q}_{\perp}+\vec{k}_{\perp})^2\Bigg[4\frac{q^-}{p^-}+4+2\left(\frac{q^-}{p^-}\right)^2\Bigg]\Theta(Qz_c-q^-)\delta^2(\vec{q}_{Tn}-\vec{q}_{\perp}-\vec{k}_{\perp})\nn
\eea
\normalsize

\subsection{Virtual Gluon emissions for $\mathcal{J}_{n,R}$}
As for the real emission, we will evaluate one diagram here in detail quoting the integrands for all the other diagrams.  We begin with diagrams in Fig.\ref{VGW}.
\bea
6(a)&=& 2g^2t_M \text{Tr}\Big[T^CT^AT^BT^C\Big] \int d^4pp^- \delta^+(p^2)\int \frac{d^dq}{q^-(2\pi)^d}\frac{i\delta^2(\vec{p}_{\perp}+\vec{k}_{\perp})\delta(Q-p^-)}{q^2-m_D^2+i\epsilon}\nn\\
&& \int \frac{dl^+}{2\pi}\frac{ie^{-it_M/2(l^+-p^+)}}{l^++i\epsilon}\frac{1}{l^+-q^+-\frac{q_{\perp}^2}{Q-q^-}+i\epsilon}\int \frac{dr^+}{2\pi}\frac{(-i)e^{-it_M/2(p^+-r^+)}}{r^+-i\epsilon}\nn\\
&&\text{Sinc}\Big[\frac{t_M}{2}(l^+-r^+)\Big]\delta^2(\vec{q}_{Tn}-\vec{k}_{\perp})
\eea  
 We can do the integral over p and then $r^+$ closing the contour in upper half plane.  Then we can do the contour integral over $q^+$,  closing the contour in the lower half of the plane.  We have two contributions 
\bea
 6(a)&=& \frac{1}{2}g^2t_M C_F\frac{\delta^{AB}}{2}\int \frac{dl^+}{2\pi}\frac{e^{-it_M/2l^+}}{l^++i\epsilon/ Q}\int \frac{d^2q_{\perp}}{(2\pi)^{d-1}}\text{Sinc}\Big[\frac{t_M}{2}(l^+)\Big]\nn\\
&&\Bigg\{\int_0^{\infty} \frac{dq^-}{(q^-)^2} \frac{1}{l^+-\frac{q_{\perp}^2+m_D^2}{q^-}-\frac{q_{\perp}^2}{Q-q^-}+\frac{i\epsilon}{Q-q^-}+\frac{i\epsilon}{q^-}}\nn\\
&-& \int_Q^{\infty} \frac{dq^-}{q^-} \frac{1}{q^-(l^+-\frac{q_{\perp}^2}{Q-q^-}+\frac{i\epsilon}{Q-q^-})-q_{\perp}^2-m_D^2+i\epsilon}\Bigg\}\nn\\
&=&\frac{1}{2}g^2t_M C_F\frac{\delta^{AB}}{2} \int \frac{dl^+}{2\pi}\frac{e^{-it_M/2l^+}}{l^++i\epsilon/ Q}\int \frac{d^2q_{\perp}}{(2\pi)^{d-1}}\text{Sinc}\Big[\frac{t_M}{2}(l^+)\Big]\nn\\
&&\int_0^{Q} \frac{dq^-}{(q^-)^2} \frac{1}{l^+-\frac{q_{\perp}^2+m_D^2}{q^-}-\frac{q_{\perp}^2}{Q-q^-}+\frac{i\epsilon}{Q-q^-}+\frac{i\epsilon}{q^-}}
\eea  
We can now do the contour integral over $l^+$ closing the contour in lower half plane  and add in the mirror diagram,  which then evaluates to 
\bea
 6(a)&=&g^2t_M C_F\frac{\delta^{AB}}{2}\delta^2(\vec{q}_{Tn}-\vec{k}_{\perp}) \int \frac{d^2q_{\perp}}{(2\pi)^{3}} \frac{1}{q_{\perp}^2+m_D^2}\Bigg\{\left(\frac{\nu}{Q}\right)^{\eta}\frac{1}{\eta}\nn\\
&-&\text{CosInt}\Big[\frac{(\vec{q}^2_{\perp}+m_D^2)t_M}{Q}\Big]
+\text{Sinc}\Big[\frac{(\vec{q}^2_{\perp}+m_D^2)t_M}{Q}\Big]\Bigg\}
\eea
  
Next consider diagram (b)   
\bea
6(b)&=& 4g^2t_M \text{Tr}\Big[T^CT^AT^CT^B\Big] \int d^4pp^- \delta^+(p^2)\int \frac{d^dq}{q^-(2\pi)^d}\frac{i\delta^2(\vec{p}_{\perp}+\vec{k}_{\perp})}{q^2-m_D^2+i\epsilon}\frac{Q-q^-}{(p-q)^2+i\epsilon}\nn\\
&& \int \frac{dl^+}{2\pi}\frac{ie^{-i\frac{t_M}{2}(l^+-p^++q^+)}}{l^+-\frac{q_{\perp}^2}{Q-q^-}+i\epsilon}\int \frac{dr^+}{2\pi}\frac{(-i)e^{-i\frac{t_M}{2}(p^+-r^+)}}{r^+-i\epsilon}\delta(Q-p^-)\nn\\
&&\text{Sinc}\Big[\frac{t_M}{2}(l^+-r^++q^+)\Big]\delta^2(\vec{q}_{Tn}-\vec{k}_{\perp})
\eea     

\bea
6(c)&=& 2g^2t_M f^{AA'C}\text{Tr}\Big[T^CT^{A'}T^B\Big] \int d^4pp^- \delta^+(p^2)\int \frac{d^dq}{q^-(2\pi)^d}\frac{i\delta(Q-q^-)\delta^2(\vec{p}_{\perp}+\vec{k}_{\perp})}{q^2-m_D^2+i\epsilon}\nn\\
&&\int \frac{dl^+}{2\pi}\frac{ie^{-i\frac{t_M}{2}(l^+-p^+)}}{l^++i\epsilon}\frac{1}{l^++q^+-\frac{q_{\perp}^2}{Q+q^-}+i\epsilon}\int \frac{dr^+}{2\pi}\frac{(-i)e^{-i\frac{t_M}{2}(p^+-r^+)}}{r^+-i\epsilon}\nn\\
&&\text{Sinc}\Big[\frac{t_M}{2}(l^+-r^+)\Big]\delta^2(\vec{q}_{Tn}-\vec{k}_{\perp})
\eea   

  \bea
   6(d)&=& 2g^2t_M f^{AA'C}\text{Tr}\Big[T^{A'}T^CT^B\Big] \int d^4pp^- \delta^+(p^2)\int \frac{d^dq}{q^-(2\pi)^d}\frac{i\delta(Q-p^-)\delta^2(\vec{p}_{\perp}+\vec{k}_{\perp})}{q^2-m_D^2+i\epsilon}\nn\\
&&\int \frac{dl^+}{2\pi}\frac{ie^{-i\frac{t_M}{2}(l^+-p^+)}}{l^++i\epsilon}\frac{1}{p^+-q^+-\frac{q_{\perp}^2}{Q-q^-}+i\epsilon}\int \frac{dr^+}{2\pi}\frac{(-i)e^{-i\frac{t_M}{2}(p^+-r^+)}}{r^+-i\epsilon}\nn\\
&&\text{Sinc}\Big[\frac{t_M}{2}(l^+-r^+)\Big]\delta^2(\vec{q}_{Tn}-\vec{k}_{\perp})
  \eea

   \bea
  6(e)&=&  2g^2t_M f^{CAD}\text{Tr}\Big[T^CT^{D}T^B\Big] \int d^4pp^- \delta^+(p^2)\int \frac{d^dq}{q^-(2\pi)^d}\frac{i2\vec{q}_{\perp}\cdot (\vec{q}_{\perp}+\vec{k}_{\perp})}{q^2-m_D^2+i\epsilon}\frac{(Q-q^-)}{(p-q)^2}\nn\\
&&\int \frac{dl^+}{2\pi}\frac{e^{-it_M/2(l^+-p^+)}}{l^++i\epsilon}\frac{1}{(l-p+q)^2+i\epsilon}\int \frac{dr^+}{2\pi}\frac{e^{-it_M/2(p^+-r^+)}}{r^+-i\epsilon}\nn\\
&&\text{Sinc}\Big[\frac{t_M}{2}(l^+-r^+)\Big]\delta^2(\vec{q}_{Tn}-\vec{k}_{\perp})\delta(Q-p^-)\delta^2(\vec{p}_{\perp}+\vec{k}_{\perp})
  \eea


\begin{thebibliography}{99}


\bibitem{Vaidya:2020lih}
V.~Vaidya,
[arXiv:2010.00028 [hep-ph]].

\bibitem{Vaidya:2021vxu}
V.~Vaidya,
[arXiv:2107.00029 [hep-ph]].

\bibitem{Gyulassy:1993hr} 
  M.~Gyulassy and X.~n.~Wang,
  Nucl.\ Phys.\ B {\bf 420}, 583 (1994)
  [nucl-th/9306003].

\bibitem{Wang:1994fx} 
  X.~N.~Wang, M.~Gyulassy and M.~Plumer,
  Phys.\ Rev.\ D {\bf 51}, 3436 (1995)
  [hep-ph/9408344].

\bibitem{Baier:1994bd} 
  R.~Baier, Y.~L.~Dokshitzer, S.~Peigne and D.~Schiff,
  Phys.\ Lett.\ B {\bf 345}, 277 (1995)
  [hep-ph/9411409].

\bibitem{Baier:1996kr} 
  R.~Baier, Y.~L.~Dokshitzer, A.~H.~Mueller, S.~Peigne and D.~Schiff,
  Nucl.\ Phys.\ B {\bf 483}, 291 (1997)
  [hep-ph/9607355].

\bibitem{Baier:1996sk} 
  R.~Baier, Y.~L.~Dokshitzer, A.~H.~Mueller, S.~Peigne and D.~Schiff,
  Nucl.\ Phys.\ B {\bf 484}, 265 (1997)
  [hep-ph/9608322].

\bibitem{Zakharov:1996fv}  
  B.~G.~Zakharov,
  JETP Lett.\  {\bf 63}, 952 (1996)
  [hep-ph/9607440].

\bibitem{Zakharov:1997uu} 
  B.~G.~Zakharov,
  JETP Lett.\  {\bf 65}, 615 (1997)
  [hep-ph/9704255].

\bibitem{Gyulassy:1999zd} 
  M.~Gyulassy, P.~Levai and I.~Vitev,
  Nucl.\ Phys.\ B {\bf 571}, 197 (2000)
  [hep-ph/9907461].

\bibitem{Gyulassy:2000er} 
  M.~Gyulassy, P.~Levai and I.~Vitev,
  Nucl.\ Phys.\ B {\bf 594}, 371 (2001)
  [nucl-th/0006010].

\bibitem{Wiedemann:2000za} 
  U.~A.~Wiedemann,
  Nucl.\ Phys.\ B {\bf 588}, 303 (2000)
  [hep-ph/0005129].

\bibitem{Guo:2000nz} 
  X.~f.~Guo and X.~N.~Wang,
  Phys.\ Rev.\ Lett.\  {\bf 85}, 3591 (2000)
  [hep-ph/0005044].
 
\bibitem{Wang:2001ifa} 
  X.~N.~Wang and X.~f.~Guo,
  Nucl.\ Phys.\ A {\bf 696}, 788 (2001)
  [hep-ph/0102230].
  
\bibitem{Arnold:2002ja} 
  P.~B.~Arnold, G.~D.~Moore and L.~G.~Yaffe,
  JHEP {\bf 0206}, 030 (2002)
  [hep-ph/0204343].

\bibitem{Arnold:2002zm} 
  P.~B.~Arnold, G.~D.~Moore and L.~G.~Yaffe,
  JHEP {\bf 0301}, 030 (2003)
  [hep-ph/0209353].
  
\bibitem{Salgado:2003gb} 
  C.~A.~Salgado and U.~A.~Wiedemann,
  Phys.\ Rev.\ D {\bf 68}, 014008 (2003)
  [hep-ph/0302184].

\bibitem{Armesto:2003jh} 
  N.~Armesto, C.~A.~Salgado and U.~A.~Wiedemann,
  Phys.\ Rev.\ D {\bf 69}, 114003 (2004)
  [hep-ph/0312106].

\bibitem{Majumder:2006wi}
A.~Majumder, B.~M\"uller and S.~A.~Bass,
Phys.\ Rev.\ Lett.\  \textbf{99}, 042301 (2007)
[arXiv:hep-ph/0611135 [hep-ph]].

\bibitem{Majumder:2007zh}
A.~Majumder, B.~M\"uller and X.~Wang,
Phys.\ Rev.\ Lett.\  \textbf{99}, 192301 (2007)
[arXiv:hep-ph/0703082 [hep-ph]].

\bibitem{Neufeld:2008fi}
R.~Neufeld, B.~M\"uller and J.~Ruppert,
Phys.\ Rev.\ C \textbf{78}, 041901 (2008)
[arXiv:0802.2254 [hep-ph]].

\bibitem{Neufeld:2009ep}
R.~Neufeld and B.~M\"uller,
Phys.\ Rev.\ Lett.\  \textbf{103}, 042301 (2009)
[arXiv:0902.2950 [nucl-th]].

\bibitem{Arsene:2004fa} 
  I.~Arsene {\it et al.} [BRAHMS Collaboration],
  Nucl.\ Phys.\ A {\bf 757}, 1 (2005)
  [nucl-ex/0410020].
  
\bibitem{Back:2004je} 
  B.~B.~Back {\it et al.},
  Nucl.\ Phys.\ A {\bf 757}, 28 (2005)
  [nucl-ex/0410022].

\bibitem{Adams:2005dq} 
  J.~Adams {\it et al.} [STAR Collaboration],
  Nucl.\ Phys.\ A {\bf 757}, 102 (2005)
  [nucl-ex/0501009].

\bibitem{Adcox:2004mh} 
  K.~Adcox {\it et al.} [PHENIX Collaboration],
  Nucl.\ Phys.\ A {\bf 757}, 184 (2005)
  [nucl-ex/0410003].

\bibitem{Aad:2010bu} 
  G.~Aad {\it et al.} [ATLAS Collaboration],
  Phys.\ Rev.\ Lett.\  {\bf 105}, 252303 (2010)
  [arXiv:1011.6182 [hep-ex]].

\bibitem{Aamodt:2010jd} 
  K.~Aamodt {\it et al.} [ALICE Collaboration],
  Phys.\ Lett.\ B {\bf 696}, 30 (2011)
  [arXiv:1012.1004 [nucl-ex]].

\bibitem{Chatrchyan:2011sx} 
  S.~Chatrchyan {\it et al.} [CMS Collaboration],
  Phys.\ Rev.\ C {\bf 84}, 024906 (2011)
  [arXiv:1102.1957 [nucl-ex]].
  
\bibitem{Mehtar-Tani:2013pia} 
  Y.~Mehtar-Tani, J.~G.~Milhano and K.~Tywoniuk,
  Int.\ J.\ Mod.\ Phys.\ A {\bf 28}, 1340013 (2013)
  [arXiv:1302.2579 [hep-ph]].

\bibitem{Blaizot:2015lma} 
  J.~P.~Blaizot and Y.~Mehtar-Tani,
  Int.\ J.\ Mod.\ Phys.\ E {\bf 24}, no. 11, 1530012 (2015)
  [arXiv:1503.05958 [hep-ph]].

\bibitem{Qin:2015srf} 
  G.~Y.~Qin and X.~N.~Wang,
  Int.\ J.\ Mod.\ Phys.\ E {\bf 24}, no. 11, 1530014 (2015)
  [arXiv:1511.00790 [hep-ph]].



\bibitem{Cao:2020wlm} 
  S.~Cao and X.~N.~Wang,
  arXiv:2002.04028 [hep-ph].

\bibitem{Bauer:2002aj}
C.~W.~Bauer, D.~Pirjol and I.~W.~Stewart,
Phys.\ Rev.\ D \textbf{67}, 071502 (2003)
[arXiv:hep-ph/0211069 [hep-ph]].

\bibitem{Bauer:2003mga}
C.~W.~Bauer, D.~Pirjol and I.~W.~Stewart,
Phys.\ Rev.\ D \textbf{68}, 034021 (2003)
[arXiv:hep-ph/0303156 [hep-ph]].


\bibitem{Manohar:2006nz}
A.~V.~Manohar and I.~W.~Stewart,
Phys.\ Rev.\ D \textbf{76}, 074002 (2007)
[arXiv:hep-ph/0605001 [hep-ph]].


\bibitem{Bauer:2000yr}
C.~W.~Bauer, S.~Fleming, D.~Pirjol and I.~W.~Stewart,
Phys.\ Rev.\ D \textbf{63}, 114020 (2001)
[arXiv:hep-ph/0011336 [hep-ph]].


\bibitem{Bauer:2001ct}
C.~W.~Bauer and I.~W.~Stewart,
Phys.\ Lett.\ B \textbf{516}, 134-142 (2001)
[arXiv:hep-ph/0107001 [hep-ph]].
  
  
\bibitem{Bauer:2002nz}
C.~W.~Bauer, S.~Fleming, D.~Pirjol, I.~Z.~Rothstein and I.~W.~Stewart,
Phys.\ Rev.\ D \textbf{66}, 014017 (2002)
[arXiv:hep-ph/0202088 [hep-ph]].

\bibitem{CaronHuot:2010bp} 
  S.~Caron-Huot and C.~Gale,
  Phys.\ Rev.\ C {\bf 82}, 064902 (2010)
  [arXiv:1006.2379 [hep-ph]].

\bibitem{Ke:2018jem}
W.~Ke, Y.~Xu and S.~A.~Bass,
Phys.\ Rev.\ C \textbf{100}, no.6, 064911 (2019)
[arXiv:1810.08177 [nucl-th]].

\bibitem{Mehtar-Tani:2019ygg} 
  Y.~Mehtar-Tani and K.~Tywoniuk,
  arXiv:1910.02032 [hep-ph].

\bibitem{Arnold:2015qya} 
  P.~Arnold and S.~Iqbal,
  JHEP {\bf 1504}, 070 (2015)
  Erratum: [JHEP {\bf 1609}, 072 (2016)]
  [arXiv:1501.04964 [hep-ph]].

\bibitem{Arnold:2016kek} 
  P.~Arnold, H.~C.~Chang and S.~Iqbal,
  JHEP {\bf 1609}, 078 (2016)
  [arXiv:1605.07624 [hep-ph]].

\bibitem{varun}
V.~Vaidya,
In preparation

\bibitem{Larkoski:2014wba}
A.~J.~Larkoski, S.~Marzani, G.~Soyez and J.~Thaler,
JHEP \textbf{05}, 146 (2014)
doi:10.1007/JHEP05(2014)146
[arXiv:1402.2657 [hep-ph]].


\bibitem{Rothstein:2016bsq}
I.~Z.~Rothstein and I.~W.~Stewart,
JHEP \textbf{08}, 025 (2016)
[arXiv:1601.04695 [hep-ph]].

\bibitem{Chiu:2012ir}
J.~Y.~Chiu, A.~Jain, D.~Neill and I.~Z.~Rothstein,
JHEP \textbf{05}, 084 (2012)
doi:10.1007/JHEP05(2012)084
[arXiv:1202.0814 [hep-ph]].


\end{thebibliography}
\end{document}